\documentclass{emulateapj}

\newcommand{\CLa}{SpARCS J161315+564930}  
\newcommand{\CLb}{SpARCS J161037+552417}  
\newcommand{\CLc}{SpARCS J161641+554513}  
\newcommand{\CLd}{SpARCS J163435+402151}  
\newcommand{\CLe}{SpARCS J163852+403843}  
\newcommand{\CLf}{SpARCS J003550-431224}  
\newcommand{\zprime}{\ensuremath{z^{\prime}}}
\newcommand{\myemail}{ricardo.demarco@ucr.edu}
\newcommand{\Mo}{\ensuremath{M_\odot}}
\newcommand{\Mstar}{\ensuremath{M_{\star}}}

\shorttitle{Spectroscopic Confirmation of Three Red-Sequence Selected
  Galaxy Clusters}

\shortauthors{Demarco et al.}

\begin{document}

\title{Spectroscopic Confirmation of Three Red-Sequence Selected
  Galaxy Clusters at $z=0.87, 1.16$ and $1.21$ from the SpARCS
  Survey\footnote{Some of the data presented herein were obtained at
    the W.M. Keck Observatory, which is operated as a scientific
    partnership among the California Institute of Technology, the
    University of California and the National Aeronautics and Space
    Administration. The Observatory was made possible by the generous
    financial support of the W.M. Keck Foundation.}}

\author{
Ricardo Demarco\altaffilmark{1,2}, 
Gillian Wilson\altaffilmark{1}, 
Adam Muzzin\altaffilmark{3}, 
Mark Lacy\altaffilmark{4}, 
Jason Surace\altaffilmark{4},
H. K. C. Yee\altaffilmark{5}, 
Henk Hoekstra\altaffilmark{6}, 
Kris Blindert\altaffilmark{7} \& 
David Gilbank\altaffilmark{8}
}

\altaffiltext{1}{Department of Physics \& Astronomy, University of
  California, Riverside. 900 University Ave. Riverside, CA 92521
  \\ \myemail} 
\altaffiltext{2}{Department of Astronomy, Universidad
  de Concepci\'on, Casilla 160-C, Concepci\'on, Chile}
\altaffiltext{3}{Department of Astronomy, Yale University. New Haven,
  CT 06520-8101} 
\altaffiltext{4}{Spitzer Science Center, California
  Institute of Technology. 220-6, Pasadena, CA 91125}
\altaffiltext{5}{Department of Astronomy \& Astrophysics, University
  of Toronto. 50 St. George St.. Toronto, Ontario M5S 3H4, Canada}
\altaffiltext{6}{Leiden Observatory, Leiden University, PO Box 9513,
  2300RA Leiden, The Netherlands} 
\altaffiltext{7}{Max-Planck-Institut
  fuer Astronomie, Koenigstuhl 17, 69117 Heidelberg, Germany}
\altaffiltext{8}{Department of Physics \& Astronomy, University of Waterloo. Waterloo, Ontario, Canada N2L 3G1}

\begin{abstract}

The Spitzer Adaptation of the Red-sequence Cluster Survey (SpARCS) is
a \zprime-passband imaging survey of the 50 deg$^2$ Spitzer SWIRE
Legacy fields, designed with the primary aim of creating the first
large, homogeneously selected sample of massive clusters at
$z~>~1$. SpARCS uses an infrared adaptation of the two-filter cluster
red-sequence technique.  In this paper we report Keck/LRIS
spectroscopic confirmation of two new exceptionally rich galaxy
clusters, \CLa\ at $z= 0.871\pm0.002$, with 14 high-confidence members
and a rest-frame velocity dispersion of $\sigma_v= 1230\pm 320$ km
s$^{-1}$, and \CLc\ at $z=1.161\pm0.003$, with seven high-confidence
members (including one AGN) and a rest-frame velocity dispersion of
$\sigma_v=950\pm 330$ km~s$^{-1}$. We also report confirmation of a
third new system, \CLb\ at $z=1.210\pm0.002$, with seven
high-confidence members and a rest-frame velocity dispersion of
$\sigma_v=410\pm300$ km s$^{-1}$. These three new spectroscopically
confirmed clusters further demonstrate the efficiency and
effectiveness of two-filter imaging for detecting \emph{bona fide}
galaxy clusters at high redshift.  We conclude by demonstrating that
prospects are good for the current generation of surveys aiming to
estimate cluster redshifts and masses at $z\gtrsim1$ \emph{directly}
from optical-infrared imaging.
\end{abstract}

\keywords{galaxies: clusters: general --- galaxies: clusters:
  individual (\CLa, \CLb, \CLc) --- galaxies: evolution --- galaxies:
  formation}

\section{Introduction}

For many years, the dominant method for discovering massive clusters
of galaxies at $z>1$, has been via X-ray emission from the hot gas in
their dark matter potential wells. Over the years, the \emph{ROSAT}
and \emph{XMM-Newton} Telescopes, in particular, have yielded a
handful of well-studied examples e.g., RDCS J0910+5422 \citep{shr02},
RDCS J1252.9-2927 \citep{rte04}, RX J0848+4452 (Lynx E;
\citealt{rse99}), XMMU J2235-25 \citep{mrl05}, and XMMXCS J2215.9-1738
\citep{srs06}. In recent years, it has been realised that, by
incorporating deep infrared (IR) observations, existing optical
imaging techniques can be adapted to successfully detect clusters at
redshifts competitive to the existing X-ray surveys. However, because
IR observations covering only modest areas (120 arcmin$^2$ - 7.25
deg$^2$) have been available, the (cluster or candidate) systems
detected to date have been less massive than those discovered from the
X-ray surveys (\citealt{seb05, bba06, vB07, mcC07, zat07, and09,
  ebg08, goto08, kri08, kurk08, mwl08}; although see \citealt{see97}).

Massive clusters of galaxies are rare, and one requires as widefield a
survey as possible to detect them.  The largest area \emph{Spitzer}
Space Telescope Survey is the 50 square degree seven passband (3.6,
4.5, 5.8, 8.0, 24, 70, 160 $\micron$) Spitzer Wide-Area Infrared
Extragalactic Survey (SWIRE) Legacy Survey \citep{lon03, sur05,
  shu08}. We have obtained deep $\zprime-$band imaging \citep{mwy09,
  wmy09}, and combined this with the pre-existing InfraRed Array
Camera (IRAC; \citealt{faz04}) observations from the SWIRE survey,
aiming to select clusters at $z > 1$ using a two filter
$\zprime-[3.6]$ infrared adaptation of the well-proven optical cluster
red sequence (RS) method \citep{gy00,gy05,gil07}.

The SpARCS\footnote{http://www.faculty.ucr.edu/$\sim$gillianw/SpARCS}
survey \citep{wmy09,mwy09}, is now complete and has an effective area
(defined as the usable overlap with SWIRE after excluding chip gaps,
regions near bright stars, etc), of 41.9 deg$^{2}$. The SpARCS catalog
contains several hundred cluster candidates at $z > 1$, by far the
largest, homogeneously selected sample of its kind.  In
\citeauthor{mwy09} and \citeauthor{wmy09} we presented an overview of
the SpARCS survey, and reported spectroscopic confirmation of three
clusters at $z=1.18, 1.20$ and $1.34$. In this paper, we report
spectroscopic confirmation of three additional clusters at $z=0.87,
1.16$ and $1.21$.

The paper is organized as follows: in \S\ref{observations}, we
describe the imaging and spectroscopic observations, and the data
reduction; in \S\ref{analysis}, we describe the spectroscopic catalog;
in \S\ref{results}, we present our results for the three individual
clusters, and estimate the cluster velocity dispersions and dynamical
masses; and in \S\ref{discussion}, we present a discussion and the
main conclusions based on our results. We assume a $\Lambda$-CDM
cosmology with $\Omega_M=0.3$ and $\Omega_{\Lambda}=0.7$, and H$_0=70$
km s$^{-1}$ Mpc$^{-1}$.

\section{Observations}\label{observations}

\subsection{Photometry}\label{photometry}

The SpARCS/SWIRE survey is comprised of six fields \citep*[see table 1
  of][]{wmy09}.  The three clusters described in this paper were
identified as high significance cluster candidates in the 7.9 deg$^2$
ELAIS-N1 field.  A full description of the SpARCS data reduction,
cluster candidate detection algorithm and catalogs will appear in
Muzzin et al.\ 2010, in preparation. We provide only a brief overview
of the main details here. SpARCS $z^{\prime}$ observations of the
Northern fields were observed by CFHT/MegaCam for a total integration
time of 6000s per pointing.  Photometry was performed on both the
$\zprime$ and IRAC mosaics using the Sextractor photometry package
\citep{ba96}.  The total 3.6 $\micron$ magnitude of all sources in the
field was computed using a large aperture, equal to the geometric mean
radius of the Sextractor isophotal aperture. Further details of the
photometric pipeline may be found in \citet{lwm05} and
\citet{mwl08}. The depth of the $z^{\prime}$ data varies from pointing
to pointing depending on the seeing and the sky background, however,
the mean 5 $\sigma$ depth for extended sources in the ELAIS-N1 field
was $\sim23.7$ Vega (24.2 AB).

The $\zprime - 3.6$ color of all sources was computed using an
aperture of diameter three IRAC pixels ($3.66 \arcsec$).  No aperture
corrections were applied to the $\zprime$ photometry because the $3.66
\arcsec$ color aperture was much larger than the median seeing.
However, the IRAC point spread function (PSF) has broad wings compared
to a typical ground-based seeing profile, and it was necessary to
apply aperture corrections to the measured 3.6 $\micron$ magnitude
before computing the color of each galaxy.

\subsection{Cluster selection}\label{selection}

Galaxy clusters were identified in the SpARCS survey, using an
algorithm very similar to that described in detail in \citet{mwl08},
as applied to the Spitzer First Look Survey (FLS;
\citealt{lwm05}). For cluster detection in both the FLS and SpARCS
surveys, we used an infrared adaptation of the cluster red-sequence
(CRS) technique \citep{gy00,gy05}. This algorithm maps the density of
galaxies in a survey within narrow color slices, giving greater weight
to brighter galaxies, and flagging the highest overdensities as
candidate clusters.  The one important difference here, compared to
\citet{mwl08}, is that the latter paper used an $R-[3.6]$ color to
detect clusters at $0 < z < 1.3$ in the FLS.  The deeper IRAC data in
SWIRE ($4\times30$s frames), compared to the FLS ($5\times12$s
frames), combined with the $\zprime-[3.6]$ color choice, allow SpARCS
to detect clusters to higher redshift than was possible with the FLS
dataset.

From analysis of the richnesses and colors, \CLc\ and \CLb\ were
identified as rich cluster candidates at $z>1$, and \CLa\ as an
unusually rich cluster candidate at $z\sim0.9$
(Table~\ref{tab_clusters}).

\subsection{Spectroscopy}

Spectroscopic observations of \CLa, \CLc\ and \CLb\ were obtained with
the Low Resolution Imaging Spectrograph (LRIS; \citealt{occ95}) on the
Keck I telescope on the nights of April 3th and April 4th, 2008. The
seeing on both nights was about $1$\arcsec. LRIS has beam-splitters
that separate the light between two sides: red \citep{occ95} and blue
\citep{mcb98}.  On the red side, we used the 400 line mm$^{-1}$
grating which gives a coverage of $\Delta \lambda = 3810$
\AA\ centered at 8500 \AA, with a dispersion of 1.86 \AA/pixels. All
the masks were designed with slits of 1\farcs1 width, which gave a
resolution of $\sim$7-9 \AA\ (FWHM) over the wavelength interval
$\sim$6600-9600 \AA. For the redshift range of these particular
clusters, all of the prominent spectral features e.g., the 4000
\AA-rest-frame break or [O$\mathrm{II}$] emission line, fall in the
LRIS red side wavelength window.

We observed one mask each in the case of clusters \CLa\ and \CLc, and
two in the case of \CLb.  The total exposure time was 6000s ($5\times
1200$) for \CLa, 8400s ($7\times 1200$) for \CLc, and 8400s ($7\times
1200$) and 7000s ($6\times 1200$) for mask $a$ and mask $b$ of
\CLb\ (Table~1).  The masks contained about 30 slits each (five of
these were alignment stars and the remaining 25 were galaxies,
selected according to a sliding scale of priority).

To reduce the number of slits placed on obvious foreground galaxies,
we prioritized slits on galaxies with colors near each cluster's
red-sequence.  To avoid significant selection bias, we used a very
broad cut around the red-sequence, intended to include both star
forming and non-star forming systems.  Slits were placed on galaxies
with priorities from 1 to 4. Priority 1 was galaxies with colors
within 0.2 magnitude of the RS, and with $[3.6] < 17.0$ (Vega).
Priority 2 was galaxies with colors between 0.2 and 0.6 magnitudes
bluer than the RS, and with $17.0 < [3.6] < 18.7$. Priority 3 was
galaxies with same colors as priority 1, but with $17.0 < [3.6] <
18.7$, and priority 4 was galaxies with same colors as priority 2, but
with $17.0 < [3.6] < 18.7$. Priorities 1 to 4 roughly correspond to
bright red-sequence, blue cloud, faint red-sequence, and faint blue
cloud galaxies respectively. Approximately ten priority 1 galaxies
could be accomodated per mask for \CLa, and five per mask for
\CLc\ and \CLb.

To reduce the LRIS data, we adapted custom software, based on that
developed by \citet{drl05,drl07} to reduce VLT/FORS2 data.  Overscan
and bias corrections were applied to both the calibration (bias, flats
and lamp arcs) and to the science (clusters and standard stars) frames
using the IRAF {\it lrisbias} and {\it lccdproc} tasks developed by
G. D. Wirth and C. Fassnacht\footnote{The LRIS software for IRAF is
  available at:
  http://www2.keck.hawaii.edu/inst/lris/kecklris.html}. After these
corrections had been applied, the data were next corrected for
geometric distortions across the dispersion axis, and then separated
into individual slitlets and reduced using standard long-slit
techniques. Each individual slitlet was processed using an algorithm
similar to that implemented in {\it bogus} (developed by D. Stern,
A. J. Bunker and S. A. Stanford)\footnote{This software can be
  obtained from:
  https://zwolfkinder.jpl.nasa.gov/$\sim$stern/homepage/bogus.html}.
The flatfield slitlets were normalized and applied to the
corresponding science slitlets. The individual 2D spectra were then
background-subtracted and fringe-corrected before being stacked to
produce a final co-added 2D spectrum.  Slits longer than 90 pixels
($\geq19$\arcsec) were further corrected for ``long-slit" distortions.
The 1D spectrum of the source (or sources) in each slitlet was then
extracted using standard IRAF tasks.

A set of NeAr lamp exposures were used to wavelength-calibrate the
observations.  LRIS spectra can suffer significant flexure which may
introduce wavelength offsets as large as $\sim16$ \AA\ in the
wavelength solution, in the most extreme cases. The offsets in the
wavelength calibration due to flexure were corrected by comparing to
the wavelengths of known skylines.  We estimate the residual
uncertainties in wavelength calibration to be at most 1 \AA, which
correspond to a redshift uncertainty of $\delta z \sim 1 \times
10^{-4}$ at $\lambda=8500$ \AA. Finally, a relative flux calibration
was performed, using a sensitivity function spanning the range 5600 -
9400 \AA\, which was obtained from long-slit observations of the
standard stars Feige 34 and Feige 67 \citep{o90}.

\section{Analysis}\label{analysis}

\subsection{Redshift catalogs}\label{z_cat}

A redshift solution was sought for each object by cross-correlating
\citep{td79} each of the LRIS spectra with observed galaxy templates
\citep{kcb96,ssp03} using the task {\it XCSAO} in the IRAF {\it RVSAO}
package \citep{kmw92}. If a redshift could be obtained for a spectrum,
it was assigned a quality flag, ``Q'', with one of two possible
values: 0 or 1. High-confidence redshifts, unambiguously determined by
two or more clear features in the continuum or in absorption, or by
obvious emission lines such as [O$\mathrm{II}$] ($\lambda3727$) were
assigned $Q=0$.  Lower-confidence redshifts based on ambiguous
identification of a few spectral features, often in low S/N spectra,
were assigned $Q=1$.

In addition to a quality flag, an emission line flag, ``E'', was also
assigned to each galaxy, based on the presence or absence of emission
line features.  Spectra displaying one or more emission feature such
as H$\beta$, [O$\mathrm{III}$]($\lambda$4959,$\lambda$5007) or
H$\alpha$ in the case of low-redshift ($z<0.8$) galaxies, or
[O$\mathrm{II}$]($\lambda$3727) in the case of higher redshift
galaxies, were assigned $E=1$.  Spectra showing no sign of excess
emission above the stellar continuum, were assigned $E=0$. The total
number of galaxies (cluster members and foreground or background
galaxies, excluding stars) extracted from the fields of \CLa,
\CLb\ and \CLc\ were 22, 19 and 34 (see Tables \ref{tab_EN1_240},
\ref{tab_EN1_349} and \ref{tab_EN1_381}).  Occasionally, one or more
serendipitous sources fell within a slit.  The breakdown by quality
flag was 20, 13 and 14 $Q=0$ galaxies, and 2, 6 and 20 $Q=1$ galaxies,
respectively.

\begin{figure}
\plotone{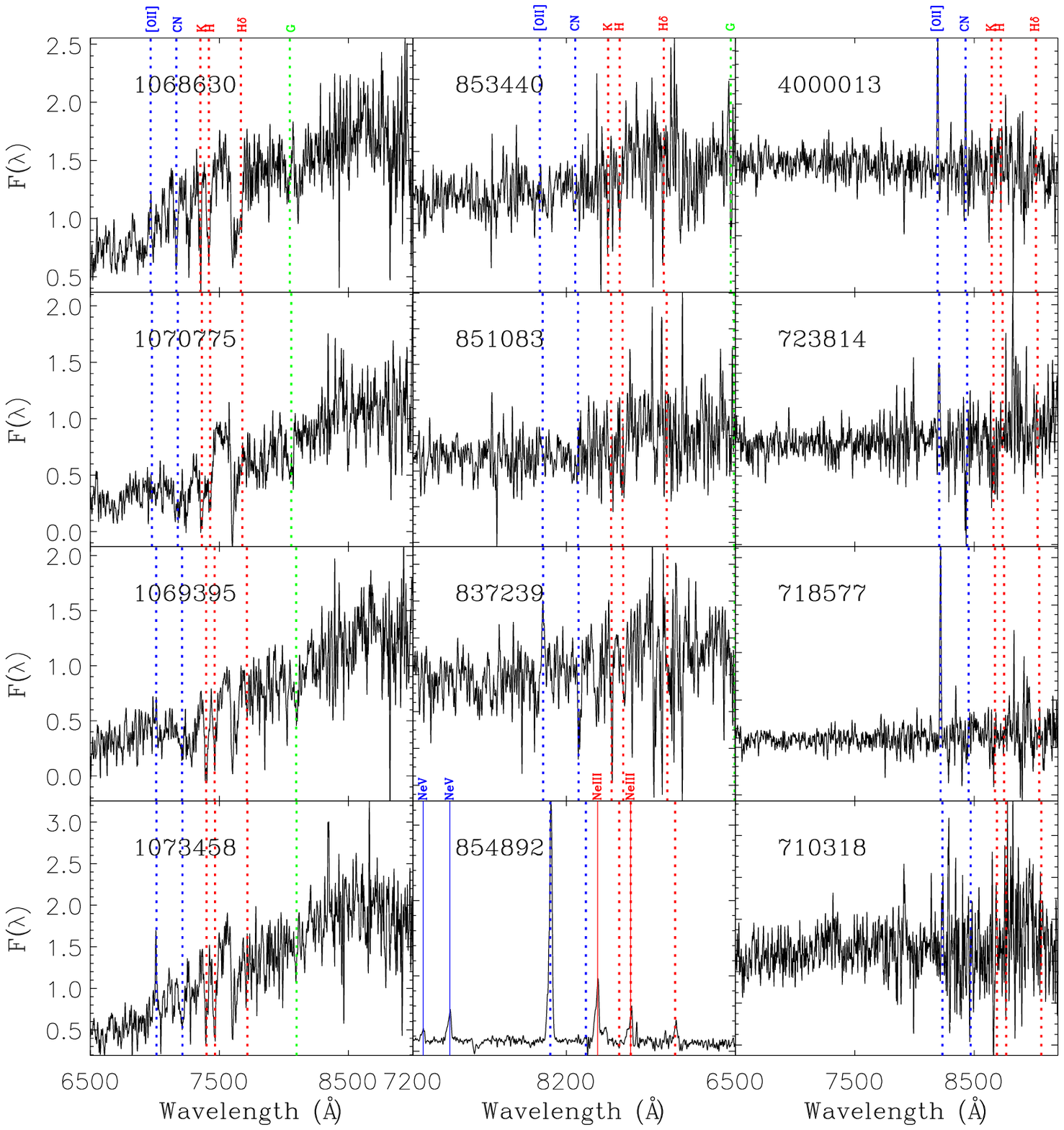}
\caption{Sample spectra of confirmed cluster members with Q=0. Left
  column shows members of cluster \CLa; center column shows members of
  \CLc; right column shows members of \CLb. Fluxes are in relative
  units, smoothed by 7 pixels (1 pix $\sim 2$ \AA). Prominent spectral
  features are indicated as vertical-dashed lines.  The lowest panel
  in the center column shows the only confirmed AGN cluster member,
  discovered in cluster \CLc\ (object 854892). The solid blue vertical
  lines indicates NeV emission lines while the solid red vertical
  lines indicate NeIII emission lines. Object 854892 also shows
  prominent [OII]($\lambda$3727) emission and high-order balmer lines
  (H$\delta$, H$\epsilon$ and H6) in emission.
\label{cl_specs} 
}  
\end{figure}

Example spectra for a subsample of cluster members (see
Section~\ref{cl240} for the definition of a ``cluster member'') are
shown in Figure~\ref{cl_specs}. The left column shows examples of
\CLa\ members, the center column shows examples of \CLc\ members, and
the right column shows exampels of \CLb\ members. The ID of each
object is indicated (Tables \ref{tab_EN1_240}, \ref{tab_EN1_349} and
\ref{tab_EN1_381}). Fluxes are in relative units, and smoothed by 7
pixels (1 pix $\sim 2$ \AA). Prominent spectral features are indicated
by vertical lines.

\section{Results}\label{results}
\label{results}

\subsection{\it \CLa}\label{cl240}

In determining cluster membership, and calculating a velocity
dispersion and mass, only high confidence, $Q=0$, galaxies were
considered.  Definitive cluster membership was determined using the
code of \citet{b06}, which is based on the shifting-gap technique of
\citet{fgg96}. This procedure uses both galaxy angular position and
radial velocity information to exclude near-field interlopers.

\begin{figure}
\plottwo{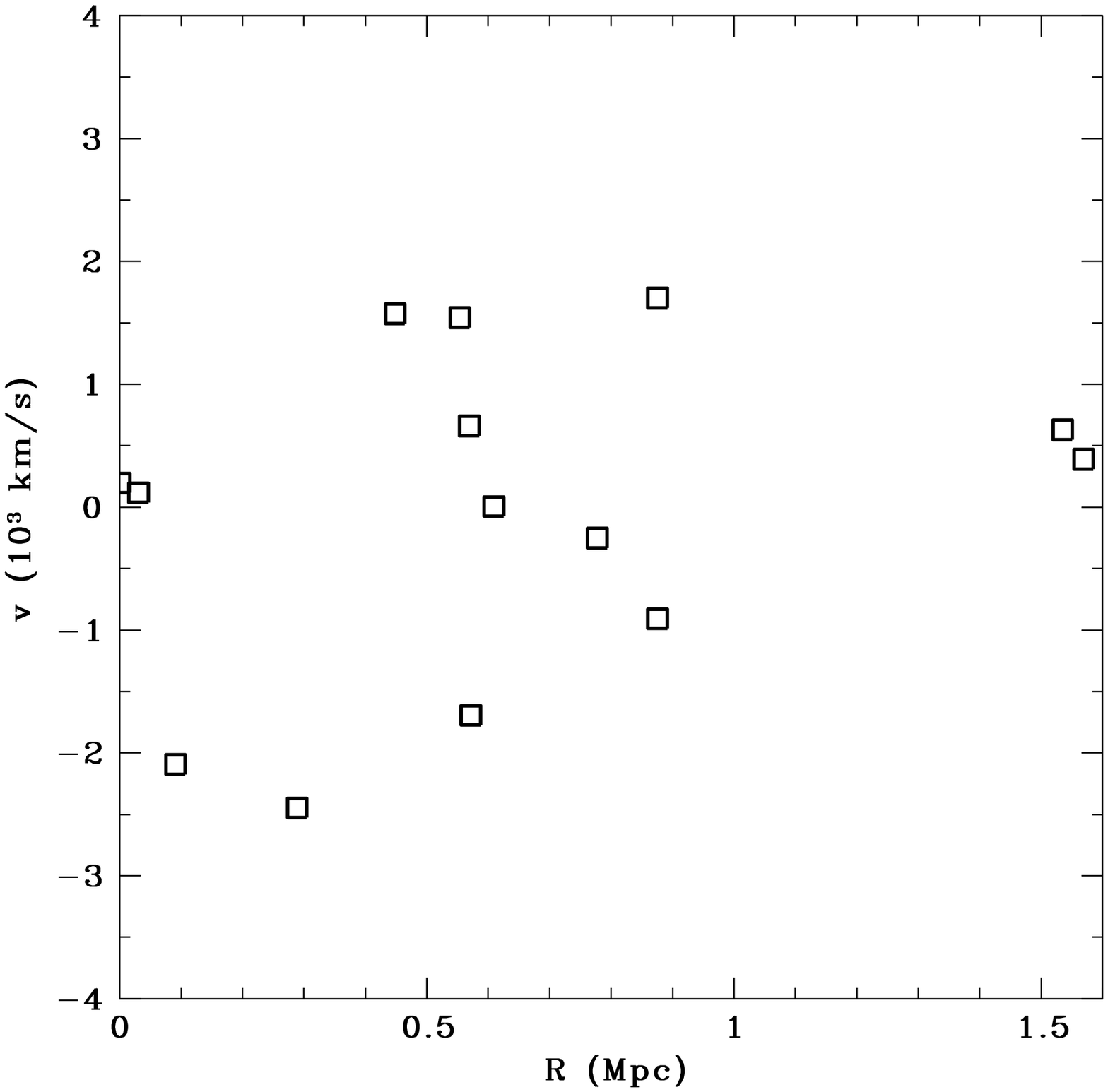}{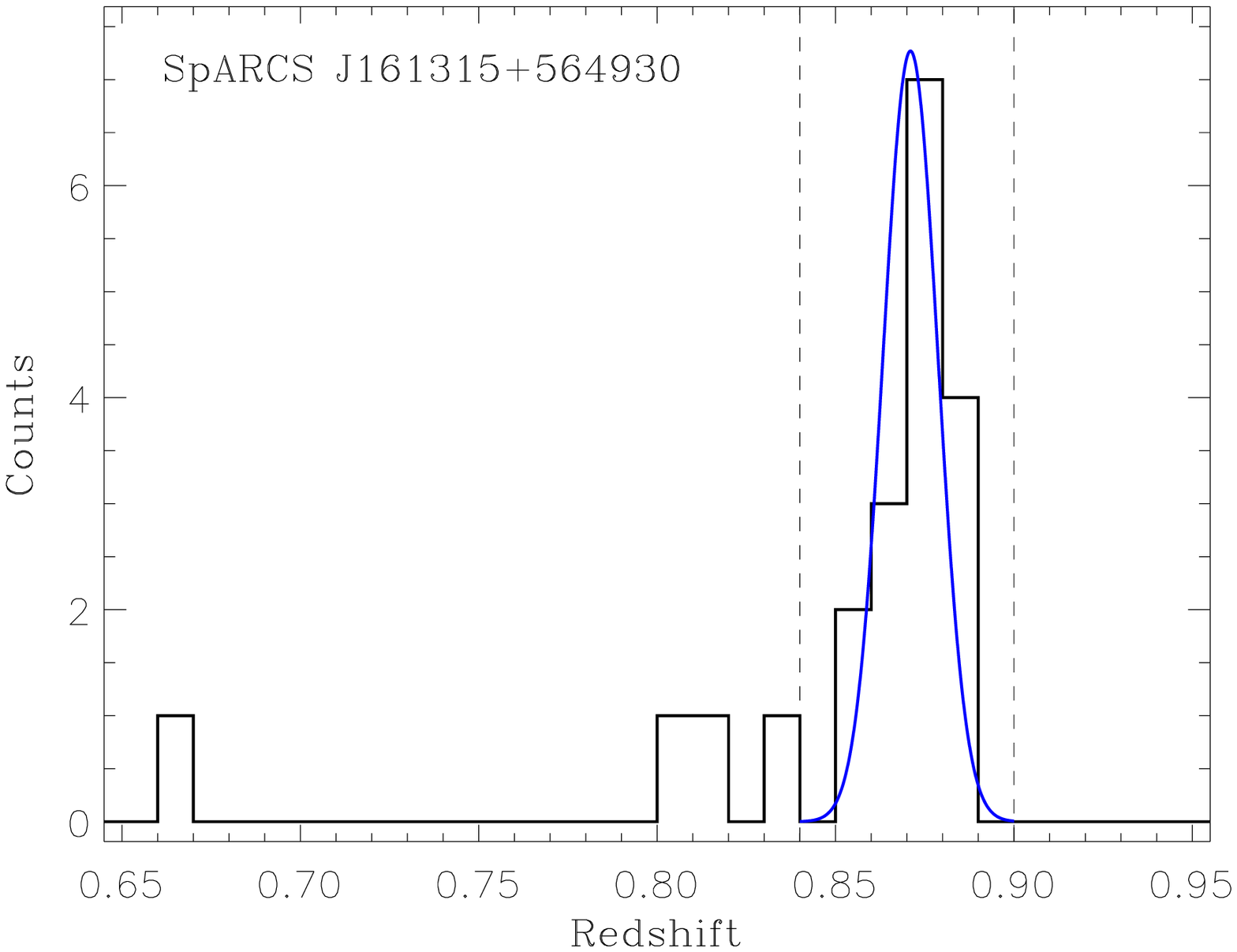}
\caption{{\it {\bf Left:}} Squares indicate galaxies with a
  high-confidence redshift (Quality flag $Q=0$), which were identified
  by the shifting-gap technique as members of cluster \CLa. Each
  galaxy velocity relative to the mean cluster velocity, is plotted as
  a function of clustercentric distance.  {\it {\bf Right:}} Histogram
  of spectroscopic redshifts in the \CLa\ field-of-view (see
  Table~\ref{tab_EN1_240}).  Confirmed members are within the redshift
  range $0.84 < z < 0.90$ indicated by the dashed vertical lines. The
  blue solid line overlaid is a Gaussian with an rms of 1230 km
  s$^{-1}$ (see Section~\ref{cl240}).
\label{z_histo_240}}
\end{figure}

The squares in the left panel of Figure~\ref{z_histo_240} show the 14
$Q=0$ galaxies identified as members of cluster \CLa\ (see
Table~\ref{tab_EN1_240}) by the shifting-gap technique.  For \CLa,
these $Q=0$ cluster members fall in the range $0.84 < z < 0.90$,
indicated by the vertical dashed lines shown in the right panel of
Figure~\ref{z_histo_240}, which shows the redshift histogram for the
\CLa\ field.

The properties of the cluster members and non-members are summarized
in Table~\ref{tab_EN1_240}. The total 3.6 $\micron$ magnitude (column
4 in Table~\ref{tab_EN1_240}), and $\zprime - 3.6$ color (column 5 in
Table~\ref{tab_EN1_240}) were calculated as described in
Section~\ref{photometry}.  In determining membership, we did not
require that any $Q=1$ galaxies satisfy the shifting-gap criteria; if
their redshifts fell in the range $0.84 < z < 0.90$ we included them
as ``cluster members" in Table~\ref{tab_EN1_240}, but re-emphasize
that we do not utilize them in estimating the cluster mean redshift or
velocity dispersion.  The total number of cluster members is 16 (14
with $Q=0$ and 2 with $Q=1$), and six foreground/background galaxies.
All confirmed members of cluster \CLa\ were passive ($E=0$) galaxies.

\begin{figure}
\plotone{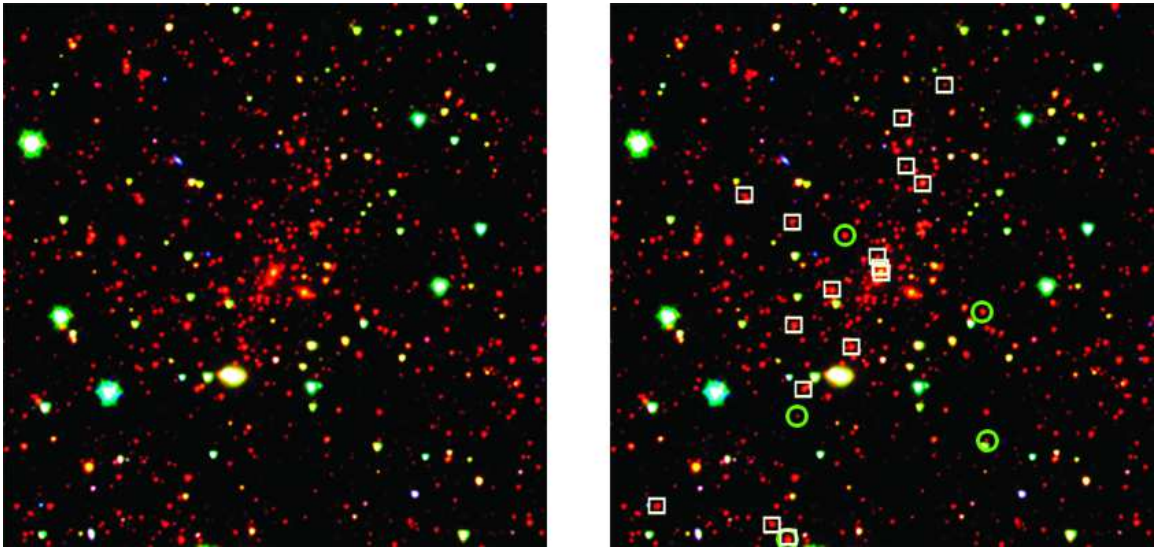}
\caption{ $r^{\prime}\zprime[3.6]$ color composites of \CLa\ are shown
  in both panels.  The FOV is $7 \arcmin$ (3.2 Mpc at the cluster
  redshift).  The white squares (green circles) overlaid on the right
  panel show the 16 $Q=0$ or $Q=1$ cluster members (and five
  foreground/background galaxies within the FOV) with
  spectroscopically-confirmed redshifts from Keck/LRIS
  (Table~\ref{tab_EN1_240}).
\label{proj_dist_240}}
\end{figure}

In some cases, a redshift was obtained which did not correspond to any
galaxy in our photometric catalog. In the case of faint galaxies, this
was because a spectroscopic redshift was obtained for a strong
emission line galaxy whose continuum fell below the detection
threshold of the $\zprime$ catalog. In the case of bright galaxies,
this was because of blending issues with the IRAC $1.8\arcsec$ $3.6
\micron$ PSF. These galaxies were assigned both a $[3.6]$-band
magnitude and a $z^{\prime}-[3.6]$ color of 99 in
Table~\ref{tab_EN1_240}. Figure~\ref{proj_dist_240} shows
$r^{\prime}z^{\prime}[3.6]$ color composites of cluster \CLa.  The
$r^{\prime}$ data were obtained from WFC on the Isaac Newton
Telescope, and are available with the SWIRE public data release
\citep{sur05}. The white squares (green circles) overlaid on the right
panel show the 16 cluster members (and foreground/background galaxies
in the $7 \arcmin$ FOV) with spectroscopically-confirmed redshifts
from Keck/LRIS (see Table~\ref{tab_EN1_240}).

Once cluster membership was established, both the redshift and
rest-frame velocity dispersion, $\sigma_z$, of SpARCS J003550-431224
were then calculated iteratively using the ``robust estimator'',
$\sigma_{rob}$ \citep{bfg90}.  The robust estimator has been shown to
be less sensitive than the standard deviation to outliers which may
persist even after rejecting interlopers using the shifting-gap
technique. The actual estimator used depends on the number of cluster
members and is either the biweight estimator for datasets with at
least 15 members, or, as here in the case of \CLa\ with 14 $Q=0$
members, the gapper estimator.  The gapper estimator is discussed more
fully in \citet{bfg90}, \citet{gbg93} and \citet{b06}.

The line-of-sight rest-frame velocity dispersion, $\sigma_v$, was
calculated directly from the vector of spectroscopic redshift
measurements, \overrightarrow{z}, as

\begin{equation}
\sigma_v=\frac{\sigma_z(\overrightarrow{z})\times c}{1+z_{cl}} \ ,
\label{veldisp}
\end{equation}

\noindent
where $c$ is the speed of light, and $\sigma_z(\overrightarrow{z})$ is
the estimated dispersion of the measured redshifts with respect to the
center of the distribution, $z_{cl}$.
 
A mean redshift of $z_{cl}=0.871\pm0.002$ and a velocity dispersion of
$\sigma_v = 1230\pm320$ km s$^{-1}$ were calculated for
\CLa\ (Table~\ref{tab_clusters}). The uncertainty on the latter was
determined using Jackknife resampling of the data. For comparison, a
Gaussian with an rms of 1230 km s$^{-1}$ has been overlaid on the
redshift histogram in the right panel of Figure~\ref{z_histo_240}.

The line-of-sight rest-frame velocity dispersion can be used to
calculate a dynamical estimate of $R_{200}(z)$, the radius at which
the mean interior density is 200 times the critical density,
$\rho_{crit}$, and $M_{200}(z)$, the mass contained within
$R_{200}(z)$.  In the spherical collapse model, at redshift $z$,
$R_{200}$ can be calculated from

\begin{equation}
R_{200} = \frac{\sqrt{3}\sigma}{10H(z)} \,
\end{equation}

\noindent
where
$H(z)=H_0\sqrt{(\Omega_M(1+z)^3+\Omega_k(1+z)^2+\Omega_{\Lambda})}$,
is the Hubble parameter at redshift $z$, and

\begin{equation}
M_{200}(z)=3 \frac{\sigma^2_v R_{200}(z)}{G} \ .
\end{equation}

\noindent
Based on its velocity dispersion of $\sigma_v = 1230\pm320$ km
s$^{-1}$, we estimate an $R_{200}= 1.9\pm0.5$ Mpc and a dynamical mass
of $M_{200}=(2.0^{+2.0}_{-1.2})\times10^{15}$ M$_{\odot}$ for
\CLa\ (Table~\ref{tab_clusters}).  Although this mass is preliminary
(see section~\ref{richness}), it seems likely that \CLa\ is an
unusually massive cluster, perhaps the most massive cluster in the
entire SpARCS survey, and comparable in mass to that of cluster MS
1054-03 at $z=0.83$ \citep{hoe00, gio04, jee05, tran07}.

\begin{figure}
\plotone{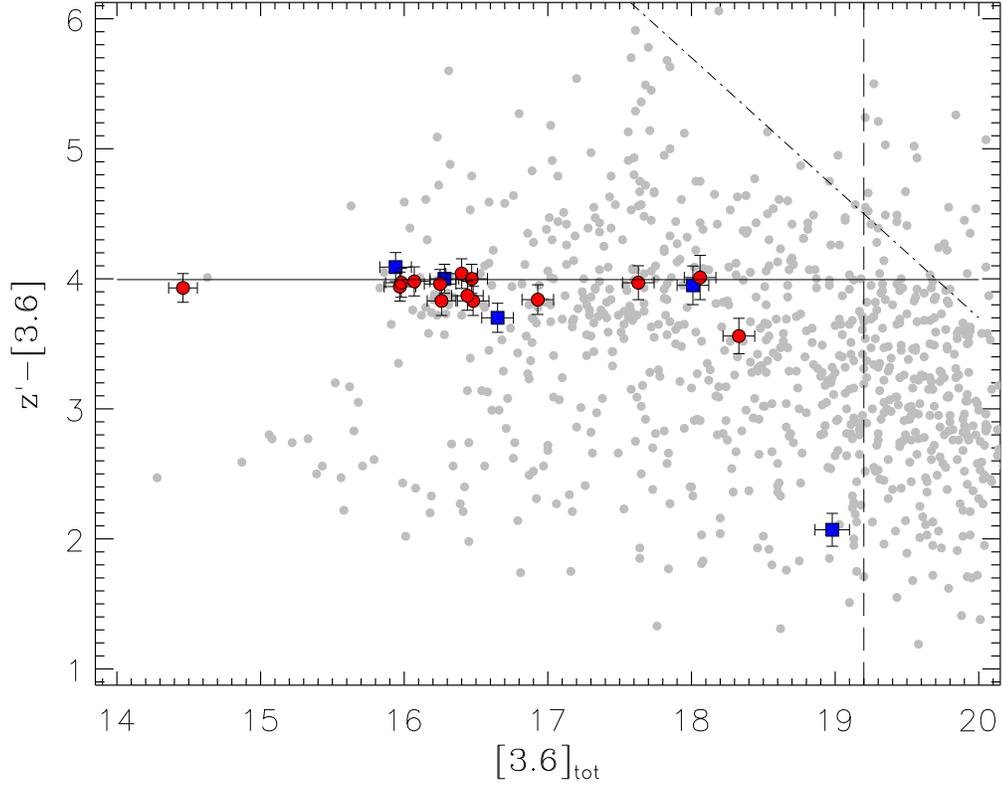}
\caption{$z^{\prime}-[3.6]$ vs. $z^{\prime}$ color-magnitude diagram
  for \CLa\ for galaxies (gray circles) which lie within a projected
  distance of $R \leq R_{200}$ ($=1.9$ Mpc) of the fiducial cluster
  center (see Table~\ref{tab_clusters}). $R_{200}$ roughly corresponds
  to the virial radius of the cluster. The colored symbols indicate
  galaxies with spectroscopic redshifts. Cluster members ($Q=0$ or
  $Q=1$) are shown by red circles and foreground/background galaxies
  by blue squares. The three galaxies in Table~\ref{tab_EN1_240}, for
  which a color could not be determined, do not appear in this figure.
  The solid line shows the predicted color of an early type galaxy at
  $z=0.87$ assuming a solar metallicity single burst model with
  formation redshift $z_{f}=4$. See~\S\ref{discussion} for a
  discussion. The vertical dashed line indicates the 5$-\sigma$ limit
  of 19.2 Vega in the 3.6 $\mu$m-band while the dot-dashed line
  indicates the color limit taking into account a 5$-\sigma$ limit in
  the $z^{\prime}$-band of 23.7 Vega.
\label{colmag_240}}
\end{figure}

\begin{figure}
\plotone{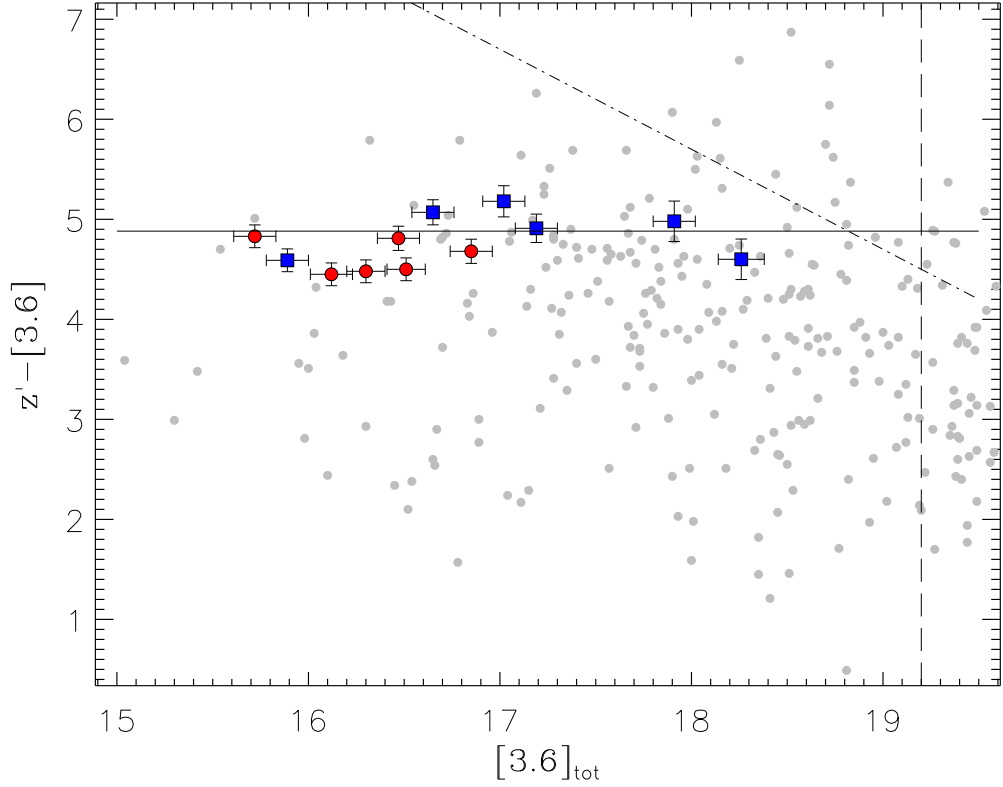}
\caption{As for Figure~\ref{colmag_240} but for \CLc, for galaxies
  which lie within a projected distance of $R \leq R_{200}$ ($=1.2$
  Mpc) of the fiducial cluster center (see Table~\ref{tab_clusters}).
  Galaxies in Table~\ref{tab_EN1_349}, which lie at $R > R_{200}$, or
  for which a color could not be determined, do not appear in this
  figure.  See Section~\ref{cl349} for more details.  The solid line
  shows the predicted color of an early type galaxy at $z=1.16$
  assuming a solar metallicity single burst model with formation
  redshift $z_{f}=4$. See~\S\ref{discussion} for a discussion. The
  vertical dashed line indicates the 5$-\sigma$ limit of 19.2 Vega in
  the 3.6 $\mu$m-band while the dot-dashed line indicates the color
  limit taking into account a 5$-\sigma$ limit in the
  $z^{\prime}$-band of 23.7 Vega.
\label{colmag_349}}
\end{figure}

\begin{figure}
\plotone{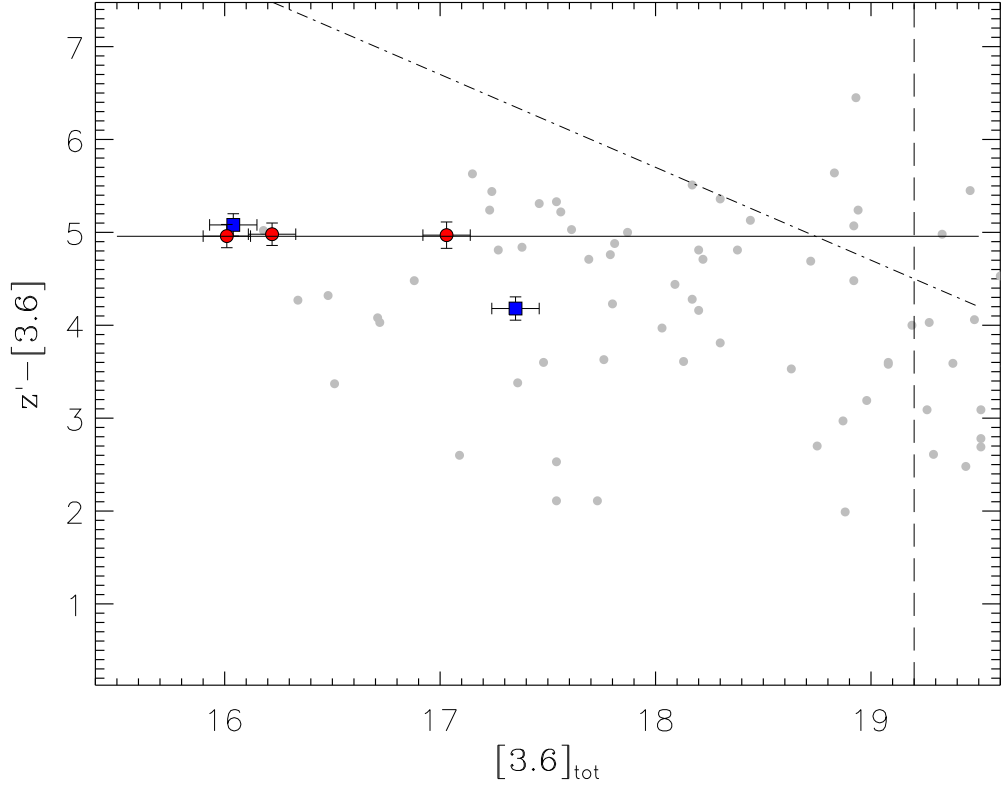}
\caption{As for Figure~\ref{colmag_240}, but for \CLb\ for galaxies at
  $R \leq R_{200}$ ($=510$ kpc) of the fiducial cluster center (see
  Table \ref{tab_clusters}).  Galaxies in Table~\ref{tab_EN1_381}, at
  $R > R_{200}$, or for which a color could not be determined, do not
  appear in this figure. The three clusters members (red circles)
  correspond to galaxies with ID $\#$'s 723814, 722784, and 722712 in
  Table~\ref{tab_EN1_381}.  See Section~\ref{cl381} for more details.
  The solid line shows the predicted color of an early type galaxy at
  $z=1.210$ assuming a solar metallicity single burst model with
  formation redshift $z_{f}=4$.  See~\S\ref{discussion} for a
  discussion. The vertical dashed line indicates the 5$-\sigma$ limit
  of 19.2 Vega in the 3.6 $\mu$m-band while the dot-dashed line
  indicates the color limit taking into account a 5$-\sigma$ limit in
  the $z^{\prime}$-band of 23.7 Vega.
\label{colmag_381}}
\end{figure}

The $z^{\prime}-[3.6]$ vs. $z^{\prime}$ color-magnitude diagram for
all galaxies (gray circles) within a radius of $R_{200}$ ($=1.9$ Mpc)
of the center of \CLa\ is shown in Figure~\ref{colmag_240}.  This
radius is approximately equal to the virial radius.
Spectroscopically-confirmed galaxies are shown by colored symbols.  In
this figure (and in Figures~\ref{colmag_349} and \ref{colmag_381}),
cluster members ($Q=0$ or $Q=1$) are shown by red circles and
foreground/background galaxies by blue squares. Note that there are
several cluster members or foreground/background galaxies shown in
Table~\ref{tab_EN1_240}, which fall within a projected radius of
$R_{200}$ of the cluster center, but for which a color could not be
determined.  These galaxies do not appear in Figure~\ref{colmag_240}
(or in Figures~\ref{colmag_349} or \ref{colmag_381}, in the cases of
Tables~\ref{tab_EN1_349} and Tables~\ref{tab_EN1_381}).

\subsection{\it \CLc}\label{cl349}

\begin{figure}
\plottwo{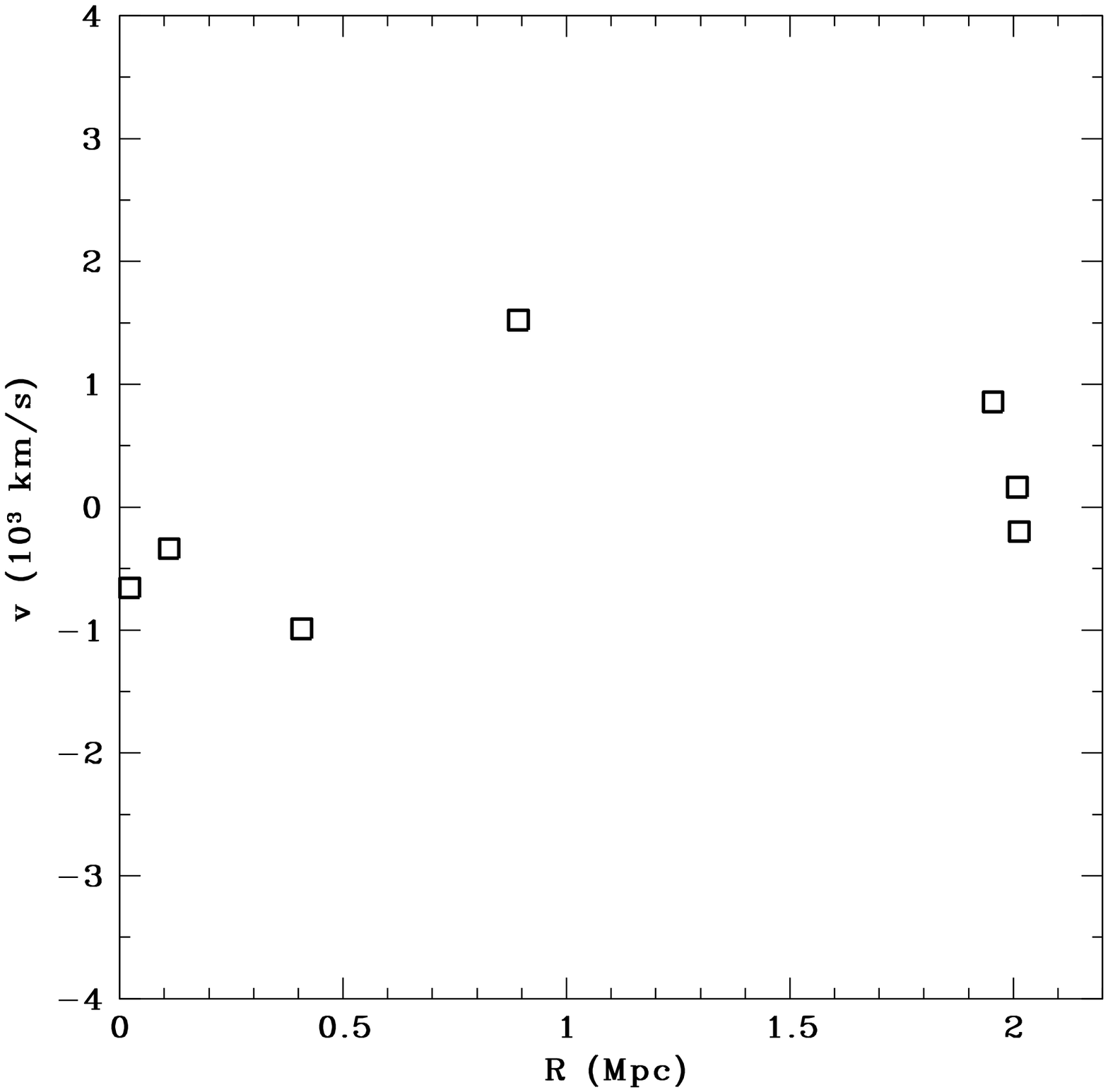}{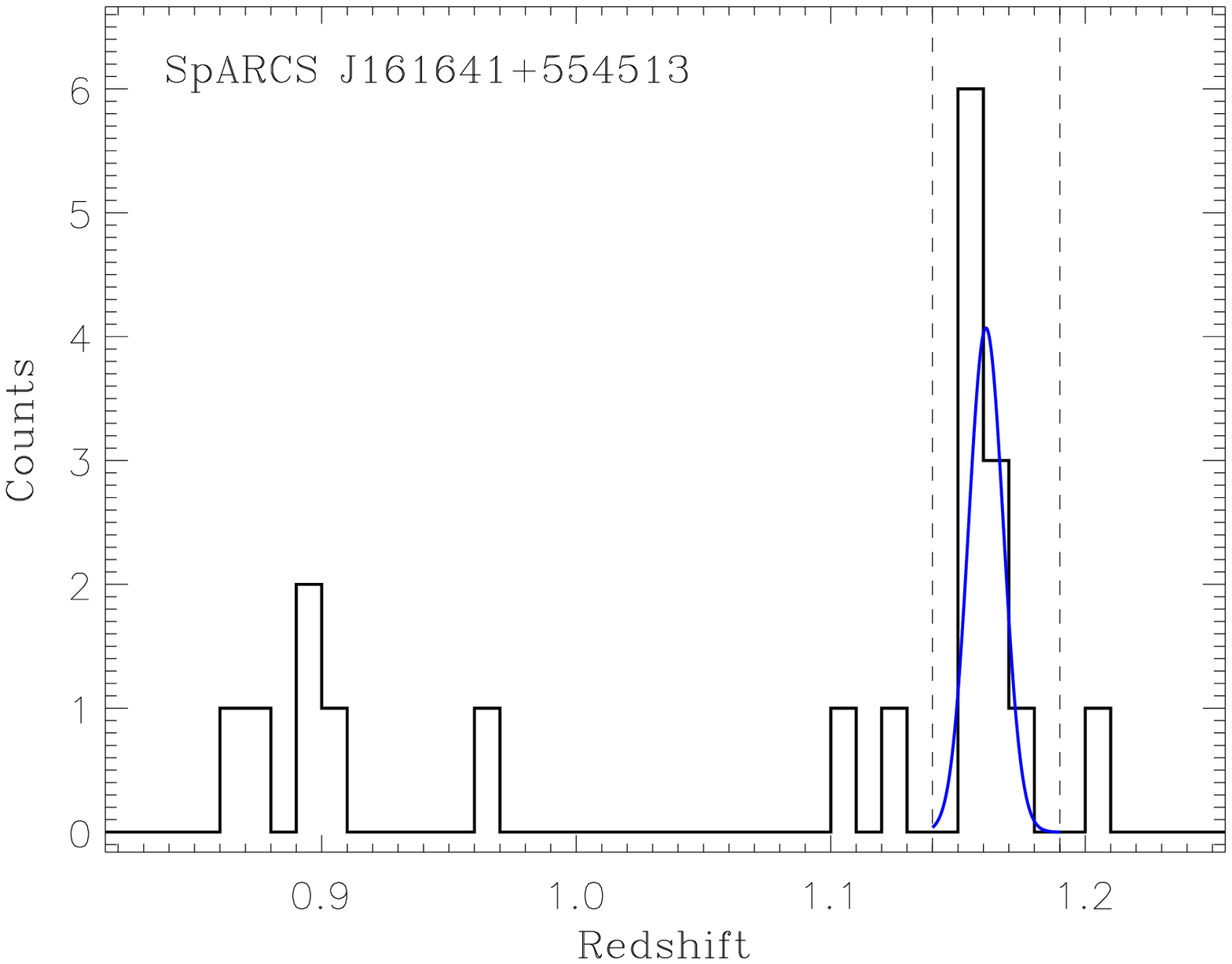}
\caption{{\it {\bf Left:}} As for Figure~\ref{z_histo_240}, but for
  the seven Q=0 galaxies identified by the shifting-gap technique as
  members of cluster \CLc\ (Table~\ref{tab_EN1_349}).  {\it {\bf
      Right:}} As for Figure~\ref{z_histo_240}, but for the
  spectroscopically confirmed galaxies in the FOV of \CLc.  Confirmed
  members are within the redshift range $1.14 < z < 1.19$ indicated by
  the dashed vertical lines.The blue solid line overlaid is a Gaussian
  with an rms of 950 km s$^{-1}$ .\label{z_histo_349}}
\end{figure}

Seven $Q=0$ galaxies were determined to be cluster members of \CLc\ by
the shifting-gap technique. These galaxies are shown by squares in the
left panel of Figure~\ref{z_histo_349}. These galaxies lie in the
redshift range $1.14 < z < 1.19$, indicated by the vertical dashed
lines in the right panel of
Figure~\ref{z_histo_349}. Table~\ref{tab_EN1_349} summarizes the
``cluster members", the ten $Q=0$ and $Q=1$ galaxies with redshifts in
this range, and the nine foreground/background galaxies.  The right
panel of Figure~\ref{z_histo_349} shows the redshift histogram for the
field of \CLc.  A Gaussian with an rms of 950 km s$^{-1}$ has been
overlaid.

\begin{figure}
\plotone{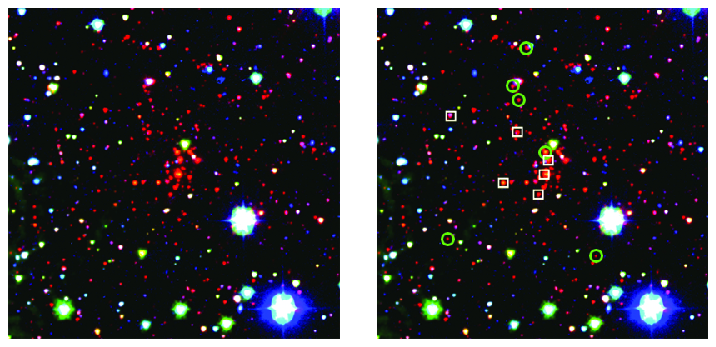}
\caption{ As for Figure~\ref{proj_dist_240}, but for \CLc.  The FOV is
  $5 \arcmin$ (2.4 Mpc at the cluster redshift).  The white squares
  (green circles) overlaid on the right panel show the cluster members
  (foreground/background galaxies) within the FOV
  (Table~\ref{tab_EN1_349}).  See Section~\ref{cl349} for a
  discussion.
\label{proj_dist_349}}
\end{figure}

Figure~\ref{proj_dist_349} shows $r^{\prime}z^{\prime}[3.6]$ color
composites of cluster \CLc.  The white squares (green circles)
overlaid on the right panel show the cluster members
(foreground/background galaxies) with spectroscopically-confirmed
redshifts (see Table~\ref{tab_EN1_349}) which fall within the
$5\times5\arcmin$ FOV of the image.

Of the seven $Q=0$ cluster members, five were classified as passive
($E=0$), one was classified as emission line ($E=1$), and one was
classifed as an AGN ($E=2$). The upper left white square in
Figure~\ref{proj_dist_349} corresponds to the AGN.  There were also
three passive $Q=1$ cluster members (Table~\ref{tab_EN1_349}). The
spectrum of the confirmed AGN (object ID $\#$ 854892 in
Table~\ref{tab_EN1_349}) is shown in the lowest panel of the center
column in Figure~\ref{cl_specs}.  This source shows
NeV($\lambda$3346,$\lambda$3426) and
NeIII($\lambda$3869,$\lambda$3968) in emission, which are common in
AGN, as well as prominent [OII]($\lambda$3727) emission and high-order
balmer lines (H$\delta$, H$\epsilon$ and H6) also in emission.

The cluster mean redshift and velocity dispersion, were estimated
iteratively from the seven $Q=0$ members, using the gapper estimator.
The mean redshift of \CLc\ was calculated to be
$z_{cl}=1.161\pm0.003$. The velocity dispersion was calculated to be
$\sigma_v = 950\pm330$ km s$^{-1}$, which corresponds to $R_{200}=
1.2\pm0.4$ Mpc, and a dynamical mass of $M_{200}=(7.7^{+11}_{-5.5})
\times10^{14}$ M$_{\odot}$(Table~\ref{tab_clusters}).

The $z^{\prime}-[3.6]$ vs. $z^{\prime}$ color-magnitude diagram for
all galaxies (gray circles) within a radius of $R_{200}$ ($=1.2$ Mpc)
of the center of \CLc\ is shown in Figure~\ref{colmag_349}.  The red
circles and blue squares indicate those cluster members and
foreground/background galaxies which lie within a projected distance
of $R_{200}$ of the cluster center, and for which a color could be
determined.

\subsection{\it \CLb}\label{cl381}

\begin{figure}
\plottwo{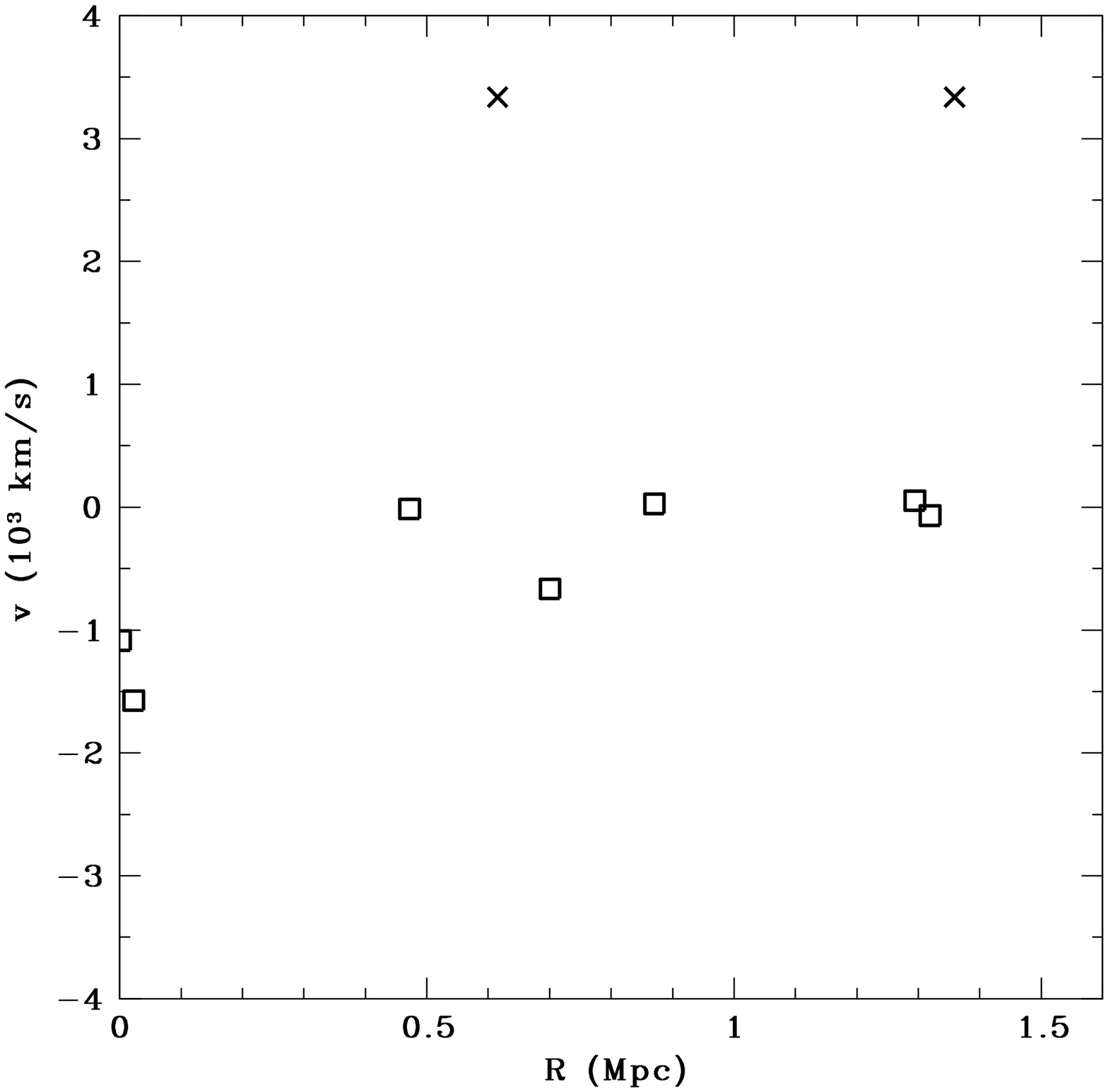}{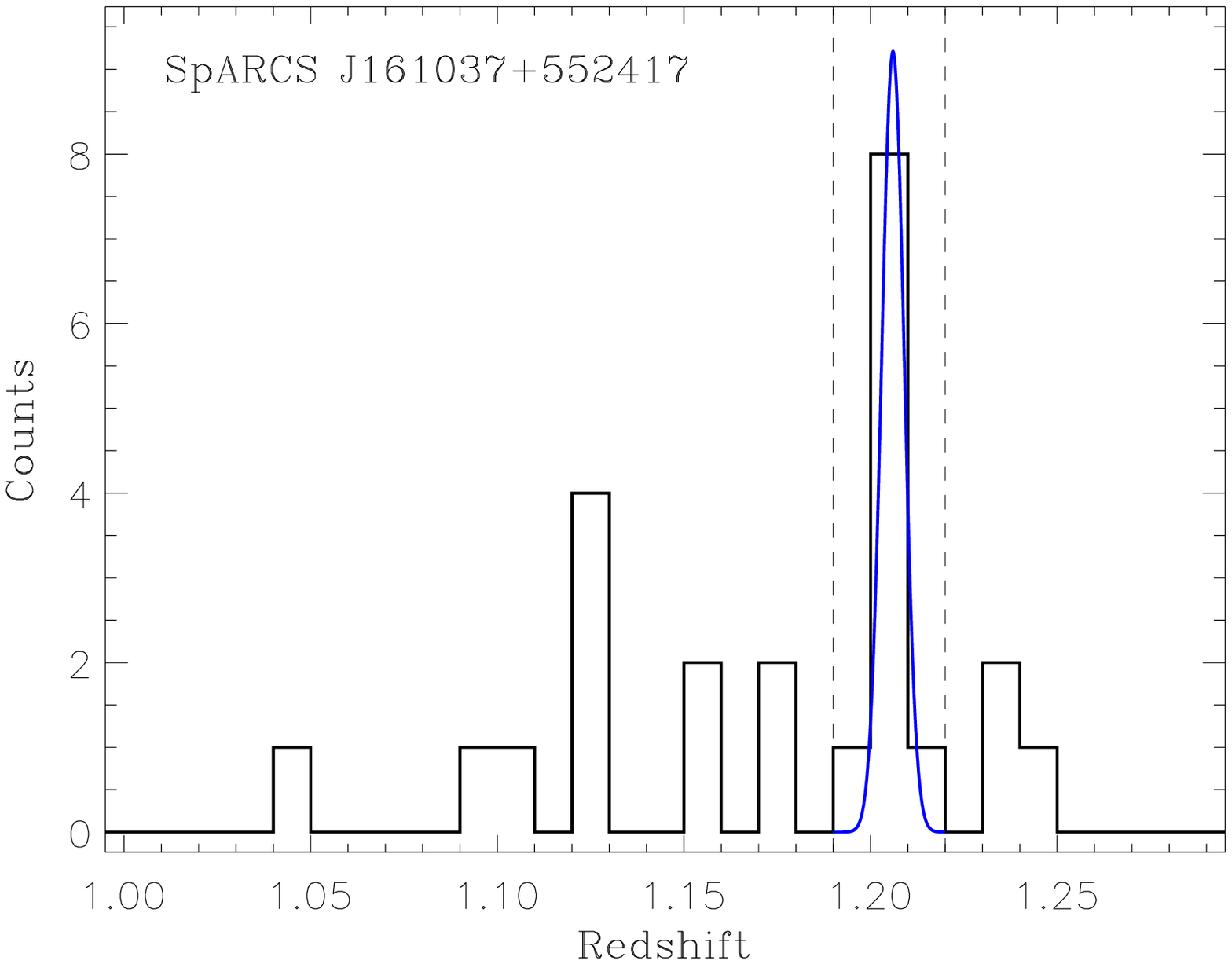}
\caption{{\it {\bf Left:}} As for Figure~\ref{z_histo_240} but for the
  seven Q=0 galaxies identified by the shifting-gap technique as
  members of cluster \CLb\ (Table~\ref{tab_EN1_381}). Two galaxies,
  shown by crosses (ID $\#$'s 727869 and 734082 in
  Table~\ref{tab_EN1_381}) were identified as near-field interlopers.
  {\it {\bf Right:}} As for Figure~\ref{z_histo_240} but for the
  spectroscopically confirmed galaxies in the FOV of \CLb.  Confirmed
  members are within the redshift range $1.19 < z < 1.22$ indicated by
  the dashed vertical lines. The blue solid line overlaid is a
  Gaussian with an rms of 410 km s$^{-1}$.
\label{z_histo_381}}
\end{figure}

Seven $Q=0$ galaxies were also determined to be cluster members of
\CLb\ by the shifting-gap technique. These galaxies are indicated by
squares in the left panel of Figure~\ref{z_histo_381}.  Two galaxies,
indicated by crosses in Figure~\ref{z_histo_381} (ID $\#$'s 727869 and
734082 in Table~\ref{tab_EN1_381}), were identified as near-field
interlopers.

The seven $Q=0$ members lie in the redshift range $1.19 < z < 1.22$,
shown by the vertical dashed lines in the right panel of
Figure~\ref{z_histo_381}. Table \ref{tab_EN1_381} summarizes the
``cluster members", the ten $Q=0$ or $Q=1$ galaxies with redshifts in
this range, and the 24 foreground/background galaxies.  The right
panel of Figure~\ref{z_histo_381} shows the redshift histogram for the
field of \CLb.  A Gaussian with an rms of 410 km s$^{-1}$ has been
overlaid.

\begin{figure}
\plotone{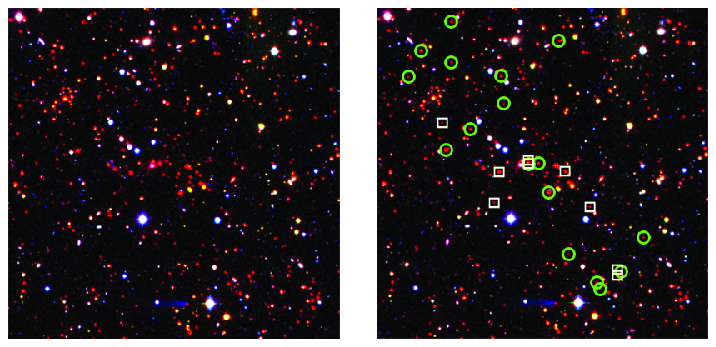}
\caption{ As for Figure~\ref{proj_dist_240}, but for \CLb. The FOV is
  6\arcmin (2.9 Mpc at the cluster redshift). The white squares (green
  circles) overlaid on the right panel show the cluster members
  (foreground/background galaxies) within the FOV
  (Table~\ref{tab_EN1_381}).  See Section~\ref{cl381} for a
  discussion.
\label{proj_dist_381}}
\end{figure}

Figure~\ref{proj_dist_381} shows $r^{\prime}z^{\prime}[3.6]$ color
composites of cluster \CLb.  The white squares (green circles)
overlaid on the right panel show the cluster members
(foreground/background galaxies) with spectroscopically-confirmed
redshifts (see Table~\ref{tab_EN1_381}) which fall within the
$6\times6 \arcmin$ FOV of the image.  Of the seven $Q=0$ cluster
members, all were classified as emission line ($E=1$).  Of the three
$Q=1$ cluster members, one was classified as passive ($E=0$), and two
as emission line ($E=1$). The cluster mean redshift and velocity
dispersion, were calculated iteratively from the seven $Q=0$ members,
using the gapper estimator.  The mean redshift of \CLb\ was estimated
to be $z_{cl}=1.210\pm0.002$ and the velocity dispersion, $\sigma_v =
410\pm300$ km s$^{-1}$, which corresponds to $R_{200}= 0.51\pm0.38$
Mpc and a dynamical mass of $M_{200}=(0.60^{+2.5}_{-0.59})
\times10^{14}$ M$_{\odot}$ (Table~\ref{tab_clusters}).

\CLb\ is significantly less massive than \CLa\ or \CLc.  The much
smaller $R_{200}$ estimated for \CLb, combined with the geometric
limitations of LRIS with respect to the redshift number density yield
measurable from a single mask, resulted in a yield of only four
spectroscopically confirmed members within $R_{200}$ from two masks.
Moreover, because of blending issues with IRAC'S $1.8\arcsec$ $3.6
\micron$ PSF, a reliable $z^{\prime}-[3.6]$ color could not be
determined for one of these galaxies (ID $\#$ 4000013 in
Table~\ref{tab_EN1_381}).  Figure~\ref{colmag_381} shows the
$z^{\prime}-[3.6]$ vs. $z^{\prime}$ color-magnitude diagram for all
galaxies (gray circles) within a radius of $R < R_{200}$ ($=510$ kpc)
of the center of cluster \CLb.  The three red circles denote the
cluster members with ID $\#$'s 723814, 722784, and 722712 in
Table~\ref{tab_EN1_381}.

\section{Discussion and Conclusions}\label{discussion}

\subsection{Red-Sequence Photometric Redshifts}\label{RS}

Color-magnitude diagrams for \CLa, \CLb\ and \CLb\ were presented in
Figures~\ref{colmag_240}, \ref{colmag_349}, and
\ref{colmag_381}. Column 2 of Table~\ref{tab_properties} shows the
$\zprime-[3.6]$ color of the red sequence for \CLa, \CLb\ and
\CLb. The color is calculated from the mean color of the
spectroscopically confirmed red-sequence cluster members. Also shown
in column 2 of Table~\ref{tab_properties} is the color of the RS for
three additional clusters, previously reported in \citet{mwy09} and
\citet{wmy09}. These clusters are \CLd\ at $z=1.180$, \CLe\ at
$z=1.196$, and \CLf\ at $z=1.335$ (column 6).  For all six clusters a
systematic uncertainty in the RS color of 0.15 magnitude has been
assumed (This uncertainty reflects the fact that, at present, we are
using the zeropoints provided by
ELIXIR\footnote{http://www.cfht.hawaii.edu/Science/CFHTLS-DATA/elixirhistory.html}
for the \zprime\ observations.  We expect, in the future, to be able
to reduce these photometric uncertainties, using our own internal
calibration).

\begin{figure}
\plotone{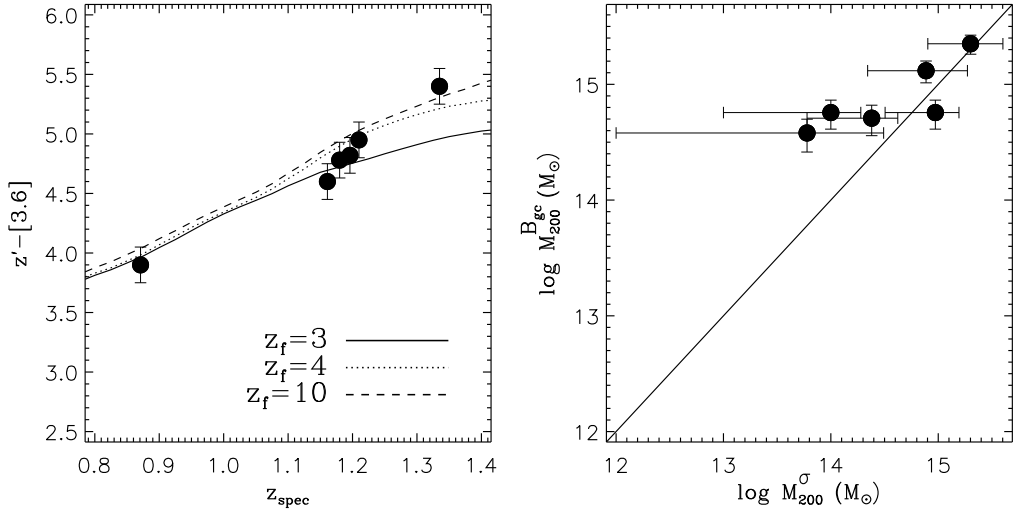}
\caption{{\it {\bf Left:}} $\zprime-[3.6]$ color vs. spectroscopic
  redshift for the six SpARCS clusters confirmed to date
  (Table~\ref{tab_properties}).  Solid, dotted and and dashed lines
  show color as a function of redshift for solar metallicity single
  burst BC03 models with formation redshifts of $z_{f}=3$, 4 and 10.
  See~\S\ref{RS} for a discussion.  {\it {\bf Right:}} Richness mass
  vs. dynamical mass estimates (see
  Table~\ref{tab_properties}). $M_{200}^{Bgc}$ is estimated from the
  $B_{gc}$ richness parameter (\S\ref{richness}), and
  $M_{200}^{\sigma}$ is estimated from the velocity dispersion
  (\S\ref{cl240} and Table~\ref{tab_clusters}).
\label{zsandmasses}}
\end{figure}

Columns 3, 4 and 5 of Table~\ref{tab_properties} show the redshift
that would be estimated for each cluster based on the measured RS
color (column 2), assuming a solar metallicity single burst BC03 model
and a formation redshift of either $z_{f}=3$, 4 or 10.  As can be seen
from Table~\ref{tab_properties} (and the left panel of
Figure~\ref{zsandmasses}), the photometric redshift inferred from the
measured $z^{\prime}-[3.6]$ color has a slight dependence on one's
choice of formation redshift, although differences in color between
the models at $z\sim1$ are fairly small ($\Delta m =0.1$ between
$z_f={4}$ and $z_f={10}$).  Utilizing the colors from the $z_{f}=4$
model, the three new clusters presented here were assigned
preliminarily redshift estimates of $z_{\rm phot}=0.84$, $z_{\rm
  phot}=1.09$, and $z_{\rm phot}=1.20$ (Table~\ref{tab_properties}).
These photometrically estimated redshifts are very similar to, albeit
slightly lower than, the spectroscopically determined values of
$z_{spec}=0.871$, $z_{spec}=1.161$, and $z_{spec}=1.210$ (column 6 in
Table~\ref{tab_properties}).

The $\zprime- [3.6]$ color vs. spectroscopic redshift for all six
SpARCS clusters in Table~\ref{tab_properties} is plotted in the left
panel of Figure~\ref{zsandmasses}. The solid, dotted and and dashed
lines show the BC03 model colors as a function of redshift for
formation redshifts of $z_{f}=3$, 4 and 10.  It is clear from
Figure~\ref{zsandmasses} that the agreement between the model colors
and the observations is very good.

With a larger sample of clusters, there may turn out to be small but
real discrepancies between the models and the measured RS colors.  In
the left panel of Figure~\ref{zsandmasses}, the $z_{f}=4$ model can be
seen to be slightly redder than the observations in the case of five
clusters (or equivalantly, the inferred photometric redshift can be
seen to be slightly lower than the spectroscopic redshift), but
slightly bluer in the case of one cluster (\CLf).  These small offsets
in color between the models and the observations, can also be seen
directly from Figures~\ref{colmag_240}, \ref{colmag_349}, and
\ref{colmag_381}. The solid lines in these three figures show the RS
color predicted by the BC03 $z_{f}=4$ model \emph{at the spectroscopic
  redshift of the cluster}, and can be seen to be slightly redder than
the observed color.  A more detailed comparison between the model
predictions and the observations will be made in a future paper
employing a larger sample of SpARCS clusters.

Despite the aforementioned caveats, and the issue of degeneracies
between the photometric redshift of the clusters and the formation
redshifts of their galaxies, our overall conclusion is that the
inferred one-color photometric redshifts and the spectroscopic
redshifts are in excellent ($\Delta z \lesssim 0.1$) general
agreement.

\subsection{Cluster Masses estimated from the Richness Parameter, $B_{gc,R}$}\label{richness}

In addition to estimating the masses for \CLa, \CLc, and \CLb\ from
the galaxy line-of-sight velocity dispersion, we also estimated the
masses from the richness of the clusters, using the $B_{gc,R}$
richness parameter.  \citet{gy05} introduced $B_{gc,R}$, an adaptation
of the $B_{gc}$ richness parameter, intended to utilize two-band
photometry to increase the contrast of the cluster with the
background, and therefore provide a measurement of the richness that
is less sensitive to foreground/background large scale structures.
$B_{gc}$ is the amplitude of the three-dimensional, cluster
center-galaxy spatial correlation function, $\xi(r) \sim B_{gc}
r^{-1.8}$ \citep{yl99}.

Instead of counting galaxies in a single passband, $B_{gc,R}$ is
obtained by counting galaxies in a color slice centered on the
location of each cluster's red-sequence in the $z^{\prime}-[3.6]$
color-magnitude diagram. In computing $B_{gc,R}$, we used a slice
bounded in color by $z^{\prime}-[3.6] = \pm0.3$ of the best-fit RS
color returned by the cluster finding algorithm (\citealt{mwl08}), and
bounded in magnitude by $(\Mstar+1)$, where $\Mstar$ is the BC03
$z_{f}=4$ model prediction of the characteristic magnitude of a galaxy
\emph{at the photometric redshift} corresponding to that RS color
(Table~\ref{tab_properties}). The background galaxy counts were
determined from the color distribution in the entire 7.9 deg$^2$
ELAIS-N1 field, minus the regions known to contain galaxy clusters.

The $B_{gc,R}$ richnesses of the three clusters were computed to be
$2452\pm422$ Mpc$^{1.8}$ (\CLa), $1762\pm358$ Mpc$^{1.8}$ (\CLc), and
$819\pm246$ Mpc$^{1.8}$ (\CLb).  Based on the empirical calibration of
$B_{\rm gc}$ vs. $M_{200}$ determined by \citet{myh07} in the K-band
for 15 CNOC1 clusters at $z \sim 0.3$, these richnesses correspond to
$M_{200} = (22.4\pm4.2) \times 10^{14} \Mo$ for \CLa, $M_{200} =
(13.1\pm2.8) \times 10^{14} \Mo$ for \CLc, and $M_{200} = (3.8\pm1.2)
\times 10^{14} \Mo$ for \CLb.

For comparison, columns 7 and 8 of Table~\ref{tab_properties} show the
dynamical mass, $M_{200}^{\sigma}$, and richness mass,
$M_{200}^{B_{gc}}$, estimates for all six SpARCS clusters.  Although
the uncertainties associated with both of the mass estimators are
large, they are consistent with each other at the 1$-\sigma$ level for
five out of the six clusters and at the 2$-\sigma$ level for the
sixth.  The agreement between the two mass estimators can be seen in
the right panel of Figure~\ref{zsandmasses}.  Based on all six SpARCS
clusters spectroscopically confirmed to date, our conclusion is that,
in addition to there being excellent agreement between the photometric
and spectroscopic redshifts, there is also reasonable agreement
between the cluster dynamical and richness mass estimates.  The
dynamical masses estimates should be considered preliminary at this
stage.  Uncertainties will reduce as more data becomes available from
a large spectroscopic follow-up program of SpARCS clusters currently
being carried out at the Gemini telescopes.

\subsection{$z>1$ Cluster Surveys}\label{surveys}

Collectively, the six SpARCS clusters confirmed to date (the three
clusters presented in \citealt{mwy09} and \citealt{wmy09}, plus the
three clusters presented here), demonstrate that, given the
availability of infrared observations, the RS technique is an
efficient and effective method of detecting \emph{bona fide} massive
galaxy clusters at $z \gtrsim 1$. Moreover, our studies of these six
clusters are showing that it is possible to infer fundamental
parameters such as cluster redshift and mass \emph{from the survey
  data itself} (see also \citealt{ebg08}).
 
At $z<1$, both the optical Red-sequence Cluster Surveys, RCS-1
\citep{gy00,gy05} and RCS-2 \citep{y07}, and The Sloan Digital Sky
Survey (SDSS; \citealt{york00, ko07b}) have shown that it is feasible
to measure cosmological parameters from the evolution of the cluster
mass function \citep{gl07, roz09b}. In order to do this efficiently,
the survey data themselves are used to detect clusters, and also to
estimate the redshift and the mass of those clusters. The redshift is
estimated from the red sequence color \citep{gil07, gl07, ko07b}, and
the mass is estimated from the optical richness \citep{ye03,
  gil07,bec07, roz09a}.  The fact that SpARCS is also now
demonstrating the practicality of estimating redshifts and masses at
$z \gtrsim 1$ from the survey data alone is heartening for the current
generation of surveys aiming to utilize optical-infrared high redshift
cluster observations to constrain cosmological parameters e.g.,
SpARCS, The UKIRT Infrared Deep-Sky Survey Deep Extragalactic Survey
(UKIDSS DXS; \citealt{law07}),and the IRAC Shallow Cluster Survey
(ISCS; \citealt{ebg08}). These optical-IR surveys will provide
complementary samples to those selected using the Sunyaev-Zel'dovich
effect, e.g., The South Pole Telescope Survey (SPT;
\citealt{ruhl04,caa09}), The Atacama Cosmology Telescope (ACT;
\citealt{kos03}), and The Atacama Pathfinder EXperiment (APEX;
\citealt{dob06}).

The complete SpARCS catalog contains several hundred cluster
candidates at $z > 1$. With new large, homogeneous, reliable $z>1$
catalogs becoming available from SpARCS and other surveys in the very
near future, the prospects look bright for high redshift cluster and
cluster galaxy evolution studies in the coming years.\\

\acknowledgments

We thank the referee for useful comments. The authors wish to
recognize and acknowledge the very significant cultural role and
reverence that the summit of Mauna Kea has always had within the
indigenous Hawaiian community.  We are most fortunate to have the
opportunity to conduct observations from this mountain. This work is
based in part on archival data obtained with the Spitzer Space
Telescope, which is operated by the Jet Propulsion Laboratory,
California Institute of Technology under a contract with NASA. Support
for this work was provided by an award issued by JPL/Caltech. GW also
gratefully acknowledges support from NSF grant AST-0909198, and from
the College of Natural and Agricultural Sciences at UCR.

{\it Facilities:} \facility{Keck (LRIS)}, \facility{Spitzer (IRAC)}, \facility{ CTIO (MOSAIC)}, \facility{CFHT (MegaCam)}.



\clearpage

\begin{deluxetable}{lcccccccccc}
\tablecolumns{11} 

\tablecaption{Summary of cluster properties. N$_{mask}$, N$_{tot}$ and
  N$_{Q=0}$ are the total number of observed masks, total number of
  spectroscopic members (Q=0 plus Q=1) and total number of
  high-confidence members only. $z_{cl}$ and $\sigma_v$ are the mean
  cluster redshift and rest-frame cluster velocity
  dispersion. $R_{200}$ is in units of Mpc, and $M_{200}$ in units of
  $\times 10^{14}$ M$_{\odot}$. $z_{cl}$, $\sigma_v$, $R_{200}$ and
  $M_{200}$ are calculated using Q=0 cluster members only.}

\tablewidth{0pt}

\tablehead{
\colhead{Cluster} & \colhead{RA (J2000)} & \colhead{DEC (J2000)} &  \colhead{N$_{mask}$} & \colhead{N$_{tot}$} & \colhead{N$_{Q=0}$} & \colhead{z$_{cl}$} & \colhead{$\sigma_v$ (km/s)} & \colhead{R$_{200}$} & \colhead{M$_{200}$} 
}
\startdata
\CLa & 16:13:14.6 & +56:49:29.9 & 1 & 16 & 14  & $0.871\pm0.002$ & $1230\pm320$ & $1.9\pm0.5$ & $20^{+20}_{-12}$ \\
\CLc & 16:16:41.3 & +55:45:12.5 & 1 & 10 & 7   & $1.161\pm0.003$ & $950\pm330$ & $1.2\pm0.4$ & $7.7^{+11}_{-5.5}$ \\
\CLb & 16:10:36.5 & +55:24:16.6 & 2 & 10 & 7   & $1.210\pm0.002$ & $410\pm300$ &$0.51\pm0.38$ & $0.60^{+2.5}_{-0.59}$ \\
\enddata
\label{tab_clusters}
\end{deluxetable}

\clearpage

\begin{deluxetable}{lccccccc}

\tablecaption{Photometric and redshift catalog for \CLa. Coordinates
  are in J2000 epoch. Column 4 is the total magnitude in the $3.6\mu$m
  band, while column 5 is the $z^{\prime}-[3.6]$ color within an
  aperture of 3\farcs66 diameter. Column 6 is the redshift and the
  uncertainty obtained from the cross correlation. Column 7 is an
  emission line flag, E; a value of 1 indicates the presence of an
  emission line in the spectrum.  Column 8 corresponds to the quality
  flag, Q, assigned to the redshift; a value of 0 is assigned to high
  confidence redshifts. Cluster members are galaxies in the range
  $0.84 < z < 0.90$. See text for details.}

\tablewidth{0pt}
\tablehead{
\colhead{ID} & \colhead{RA (deg)} & \colhead{DEC (deg)} & \colhead{[3.6]$_{\rm tot}$} & \colhead{\zprime - [3.6]} & \colhead{z} & \colhead{E} & \colhead{Q}
}
 
\startdata
\cutinhead{Cluster members}
1070485 & 243.32851 & 56.821579 & $15.98\pm0.11$ & $3.97\pm0.11$ & $0.8555\pm0.0004$ & 0 & 0 \\
1071085 & 243.31200 & 56.828140 & $17.63\pm0.11$ & $3.97\pm0.13$ & $0.8577\pm0.0006$ & 0 & 0 \\
1071707 & 243.34280 & 56.835220 & $16.47\pm0.11$ & $4.00\pm0.11$ & $0.8602\pm0.0003$ & 0 & 0 \\
1072162 & 243.36000 & 56.840450 & $16.93\pm0.11$ & $3.84\pm0.11$ & $0.8651\pm0.0003$ & 0 & 0 \\
1068630 & 243.33870 & 56.802010 & $16.26\pm0.10$ & $3.83\pm0.11$ & $0.8692\pm0.0004$ & 0 & 0 \\
1072641 & 243.30150 & 56.845970 & $18.06\pm0.11$ & $4.01\pm0.17$ & $0.8708\pm0.0004$ & 0 & 0 \\
2000012 & 243.31107 & 56.826047 & $99$ & $99$ & $0.8715\pm0.0003$ & 0 & 0 \\
1070775 & 243.31090 & 56.824970 & $14.46\pm0.10$ & $3.93\pm0.11$ & $0.8720\pm0.0003$ & 0 & 0 \\
1066451 & 243.39220 & 56.778770 & $16.44\pm0.11$ & $3.87\pm0.11$ & $0.8730\pm0.0003$ & 0 & 1 \\
2000007 & 243.34483 & 56.772531 & $99$ & $99$ & $0.8732\pm0.0004$ & 0 & 0 \\
1066059 & 243.35040 & 56.775040 & $16.40\pm0.11$ & $4.04\pm0.11$ & $0.8747\pm0.0004$ & 0 & 0 \\
1069805 & 243.34230 & 56.814430 & $16.07\pm0.11$ & $3.98\pm0.11$ & $0.8749\pm0.0004$ & 0 & 0 \\
1074041 & 243.28720 & 56.862221 & $18.33\pm0.11$ & $3.56\pm0.14$ & $0.8801\pm0.0006$ & 0 & 1 \\
1072364 & 243.29550 & 56.842720 & $15.97\pm0.11$ & $3.94\pm0.11$ & $0.8804\pm0.0003$ & 0 & 0 \\
1069395 & 243.32159 & 56.810188 & $16.25\pm0.11$ & $3.96\pm0.11$ & $0.8806\pm0.0003$ & 0 & 0 \\
1073458 & 243.30260 & 56.855709 & $16.48\pm0.11$ & $3.83\pm0.11$ & $0.8814\pm0.0002$ & 0 & 0 \\
\cutinhead{Foreground/background galaxies}
1068289 & 243.33870 & 56.797932 & $18.98\pm0.12$ & $2.07\pm0.13$ & $0.3551\pm0.0001$ & 1 & 0 \\
1071481 & 243.32381 & 56.832401 & $16.28\pm0.10$ & $4.00\pm0.11$ & $0.6657\pm0.0002$ & 0 & 0 \\
1070036 & 243.27380 & 56.817139 & $16.65\pm0.11$ & $3.70\pm0.11$ & $0.8018\pm0.0004$ & 0 & 0 \\
1067689 & 243.27271 & 56.791660 & $18.01\pm0.11$ & $3.95\pm0.15$ & $0.8163\pm0.0002$ & 1 & 0 \\
1065709 & 243.34480 & 56.771900 & $15.94\pm0.11$ & $4.09\pm0.11$ & $0.8393\pm0.0002$ & 1 & 0 \\
2000006 & 243.34517 & 56.771469 & $99$ & $99$ & $1.1701\pm0.0002$ & 1 & 0 \\
\enddata
\label{tab_EN1_240}
\end{deluxetable}

\clearpage

\begin{deluxetable}{lccccccc}

\tablecaption{As for Table~\ref{tab_EN1_240}, but for \CLc. The AGN
  has been assigned a value of $E=2$ }

\tablewidth{0pt}
\tablehead{
\colhead{ID} & \colhead{RA (deg)} & \colhead{DEC (deg)} & \colhead{[3.6]$_{\rm tot}$} & \colhead{\zprime - [3.6]} & \colhead{z} & \colhead{E} & \colhead{Q}
}
 
\startdata
\cutinhead{Cluster members}
848097 & 244.17630 & 55.748390 & $16.85\pm0.11$ & $4.68\pm0.12$ & $1.1506\pm0.0003$ & 0 & 1 \\
853440 & 244.18610 & 55.764381 & $16.47\pm0.11$ & $4.81\pm0.12$ & $1.1537\pm0.0003$ & 0 & 0 \\
849838 & 244.17340 & 55.753590 & $16.51\pm0.10$ & $4.50\pm0.11$ & $1.1561\pm0.0003$ & 0 & 0 \\
849110 & 244.19260 & 55.751389 & $16.12\pm0.11$ & $4.45\pm0.11$ & $1.1562\pm0.0007$ & 0 & 1 \\
851083 & 244.17180 & 55.757141 & $15.72\pm0.11$ & $4.83\pm0.11$ & $1.1584\pm0.0003$ & 0 & 0 \\
837239 & 244.26920 & 55.715519 & $16.15\pm0.11$ & $4.59\pm0.11$ & $1.1594\pm0.0005$ & 1 & 0 \\
846319 & 244.25529 & 55.743149 & $16.41\pm0.10$ & $4.62\pm0.12$ & $1.1605\pm0.0008$ & 0 & 1 \\
3000009 & 244.26935 & 55.715914 & $99$ & $99$ & $1.1620\pm0.0003$ & 0 & 0 \\
839226 & 244.27161 & 55.721340 & $16.88\pm0.11$ & $4.64\pm0.12$ & $1.1670\pm0.0005$ & 0 & 0 \\
854892 & 244.21670 & 55.768921 & $16.30\pm0.10$ & $4.48\pm0.11$ & $1.1718\pm0.0001$ & 2 & 0 \\
\cutinhead{Foreground/background galaxies}
833647 & 244.16330 & 55.704411 & $17.64\pm0.11$ & $5.10\pm0.19$ & $0.8673\pm0.0004$ & 1 & 0 \\
828383 & 244.22971 & 55.688580 & $17.37\pm0.11$ & $4.46\pm0.13$ & $0.8790\pm0.0002$ & 1 & 0 \\
840274 & 244.26221 & 55.724548 & $17.61\pm0.11$ & $4.16\pm0.13$ & $0.8905\pm0.0005$ & 1 & 0 \\
856147 & 244.18530 & 55.772751 & $17.19\pm0.11$ & $4.91\pm0.14$ & $0.8996\pm0.0002$ & 1 & 0 \\
844301 & 244.21809 & 55.736950 & $17.91\pm0.11$ & $4.98\pm0.20$ & $0.9026\pm0.0004$ & 1 & 0 \\
842789 & 244.15030 & 55.732430 & $18.26\pm0.12$ & $4.60\pm0.20$ & $0.9629\pm0.0005$ & 0 & 1 \\
857305 & 244.18820 & 55.776299 & $17.02\pm0.11$ & $5.18\pm0.16$ & $1.1096\pm0.0005$ & 0 & 0 \\
851794 & 244.17329 & 55.759270 & $15.89\pm0.11$ & $4.59\pm0.11$ & $1.1242\pm0.0004$ & 0 & 1 \\
860365 & 244.18150 & 55.786140 & $16.65\pm0.11$ & $5.07\pm0.13$ & $1.2041\pm0.0007$ & 0 & 1 \\
\enddata
\label{tab_EN1_349}
\end{deluxetable}

\clearpage

\begin{deluxetable}{lccccccc}

\tablecaption{As for Table~\ref{tab_EN1_240}. but for \CLb.}

\tablewidth{0pt}
\tablehead{
\colhead{ID} & \colhead{RA (deg)} & \colhead{DEC (deg)} & \colhead{$[3.6]_{\rm tot}$} & \colhead{$\zprime - [3.6]$} & \colhead{z} & \colhead{E} & \colhead{Q}
}

\startdata
\cutinhead{Cluster members}
4000013 & 242.65226 & 55.405353 & $99$ & $99$ & $1.1981\pm0.0002$ & 1 & 0 \\
723814 & 242.65221 & 55.404598 & $16.01\pm0.11$ & $4.96\pm0.13$ & $1.2017\pm0.0003$ & 1 & 0 \\
722784 & 242.63251 & 55.401890 & $17.03\pm0.11$ & $4.97\pm0.14$ & $1.2047\pm0.0006$ & 0 & 1 \\
718577 & 242.61940 & 55.391060 & $18.16\pm0.12$ & $4.21\pm0.17$ & $1.2048\pm0.0001$ & 1 & 0 \\
722712 & 242.66769 & 55.401711 & $16.22\pm0.11$ & $4.98\pm0.12$ & $1.2058\pm0.0005$ & 1 & 1 \\
4000026 & 242.60676 & 55.475206 & $99$ & $99$ & $1.2089\pm0.0004$ & 1 & 1 \\
710318 & 242.60490 & 55.370651 & $17.00\pm0.11$ & $5.55\pm0.22$ & $1.2092\pm0.0003$ & 1 & 0 \\
4000010 & 242.67026 & 55.393053 & $99$ & $99$ & $1.2096\pm0.0004$ & 1 & 0 \\
728525 & 242.69780 & 55.416550 & $18.32\pm0.12$ & $4.62\pm0.22$ & $1.2099\pm0.0006$ & 1 & 0 \\
4000032 & 242.60455 & 55.371836 & $99$ & $99$ & $1.2101\pm0.0019$ & 1 & 0 \\
\cutinhead{Foreground/background galaxies}
744358 & 242.57840 & 55.457119 & $17.28\pm0.11$ & $5.67\pm0.22$ & $0.7646\pm0.0004$ & 1 & 0 \\
4000012 & 242.65180 & 55.404111 & $99$ & $99$ & $0.8118\pm0.0003$ & 1 & 0 \\
723736 & 242.64661 & 55.404339 & $17.35\pm0.11$ & $4.18\pm0.13$ & $0.9399\pm0.0008$ & 0 & 1 \\
709346 & 242.61549 & 55.368011 & $16.83\pm0.11$ & $5.14\pm0.16$ & $0.9615\pm0.0004$ & 0 & 1 \\
714998 & 242.59080 & 55.381821 & $17.89\pm0.11$ & $4.00\pm0.13$ & $0.9667\pm0.0009$ & 1 & 1 \\
710721 & 242.60280 & 55.371490 & $16.96\pm0.11$ & $5.05\pm0.15$ & $0.9803\pm0.0004$ & 0 & 0 \\
751852 & 242.60730 & 55.476471 & $17.56\pm0.11$ & $5.09\pm0.25$ & $0.9982\pm0.0009$ & 0 & 1 \\
734249 & 242.66650 & 55.430779 & $17.06\pm0.11$ & $4.66\pm0.14$ & $1.0435\pm0.0006$ & 0 & 1 \\
747661 & 242.62151 & 55.465469 & $17.99\pm0.11$ & $4.64\pm0.20$ & $1.0948\pm0.0005$ & 1 & 1 \\
740510 & 242.69310 & 55.447041 & $17.19\pm0.11$ & $5.66\pm0.28$ & $1.1060\pm0.0005$ & 0 & 1 \\
708779 & 242.61410 & 55.366550 & $18.11\pm0.11$ & $4.92\pm0.30$ & $1.1201\pm0.0009$ & 1 & 1 \\
725337 & 242.69600 & 55.408329 & $16.10\pm0.11$ & $5.10\pm0.12$ & $1.1255\pm0.0006$ & 0 & 1 \\
4000015 & 242.60952 & 55.453100 & $99$ & $99$ & $1.1265\pm0.0005$ & 0 & 1 \\
730800 & 242.66510 & 55.422260 & $18.29\pm0.12$ & $4.98\pm0.31$ & $1.1274\pm0.0006$ & 0 & 0 \\
720317 & 242.64140 & 55.395630 & $16.04\pm0.11$ & $5.08\pm0.12$ & $1.1559\pm0.0005$ & 0 & 1 \\
745152 & 242.66029 & 55.459110 & $17.08\pm0.11$ & $5.53\pm0.22$ & $1.1565\pm0.0010$ & 0 & 1 \\
753137 & 242.63040 & 55.479912 & $16.53\pm0.11$ & $4.71\pm0.12$ & $1.1753\pm0.0007$ & 0 & 1 \\
738292 & 242.63560 & 55.441189 & $18.15\pm0.11$ & $4.41\pm0.18$ & $1.1772\pm0.0006$ & 0 & 0 \\
727869 & 242.68291 & 55.414661 & $17.08\pm0.11$ & $4.63\pm0.13$ & $1.2343\pm0.0003$ & 1 & 0 \\
734082 & 242.71609 & 55.430450 & $18.18\pm0.11$ & $4.98\pm0.25$ & $1.2343\pm0.0003$ & 1 & 0 \\
712763 & 242.63080 & 55.376591 & $18.04\pm0.11$ & $5.21\pm0.30$ & $1.2404\pm0.0004$ & 1 & 1 \\
742510 & 242.67841 & 55.452202 & $17.24\pm0.11$ & $5.39\pm0.19$ & $1.4546\pm0.0005$ & 1 & 1 \\
737166 & 242.70931 & 55.438370 & $16.67\pm0.10$ & $5.29\pm0.14$ & $1.4789\pm0.0003$ & 1 & 1 \\
735789 & 242.69299 & 55.434792 & $17.48\pm0.11$ & $5.52\pm0.25$ & $1.4799\pm0.0003$ & 1 & 1 \\
\enddata
\label{tab_EN1_381}
\end{deluxetable}

\clearpage

\begin{deluxetable}{lccccccc}

\tablecaption{Summary of redshifts and mass estimates for the six
  SpARCS clusters confirmed to date.  Columns show measured $\zprime -
  [3.6]$ color, redshift corresponding to color assuming solar
  metallicity single burst BC03 model with formation redshift
  $z_{f}=3$, 4 or 10, spectroscopic redshift~(see
  Table~\ref{tab_clusters}), $M_{200}$ estimated from velocity
  dispersion~(see Table~\ref{tab_clusters}), and $M_{200}$ estimated
  from richness.  The last two columns are in units of $\times
  10^{14}$ M$_{\odot}$.  }

\tablewidth{0pt}

\tablehead{
\colhead{Cluster} & \colhead{$\zprime - [3.6]$ color}  & \colhead{z($z_f=3$)} & \colhead{z($z_f=4$)}  & \colhead{z($z_f=10$)}  & \colhead{z$_{cl}$} & \colhead{M$_{200}^{\sigma}$} & \colhead{$M_{200}^{Bgc}$}  
}
\startdata
\CLa & $3.90\pm0.15$  & $0.85^{+0.05}_{-0.07}$ & $0.84^{+0.06}_{-0.08}$  & $0.81^{+0.07}_{-0.07}$ & $0.871\pm0.002$ & $20^{+20}_{-12}$      & $22.4\pm4.2$   \\
\CLc & $4.60\pm0.15$  & $1.11^{+0.09}_{-0.06}$ & $1.09^{+0.05}_{-0.04}$  & $1.08^{+0.05}_{-0.05}$ & $1.161\pm0.003$ & $7.7^{+11}_{-5.5}$    & $13.1\pm2.8$  \\
\CLb & $4.95\pm0.15$  & $1.32^{+0.23}_{-0.09}$ & $1.20^{+0.07}_{-0.05}$  & $1.18^{+0.06}_{-0.05}$ & $1.210\pm0.002$ & $0.60^{+2.5}_{-0.59}$ & $3.8\pm1.2$ \\
\CLd\tablenotemark{a} & $4.77\pm0.15$  & $1.21^{+0.10}_{-0.09}$ & $1.14^{+0.05}_{-0.03}$  & $1.13^{+0.04}_{-0.04}$ & $1.180\pm0.002$ & $1.0\pm0.9$           & $5.7\pm1.6$ \\
\CLe\tablenotemark{a} & $4.82\pm0.15$  & $1.24^{+0.09}_{-0.09}$ & $1.16^{+0.05}_{-0.04}$  & $1.14^{+0.05}_{-0.04}$ & $1.196\pm0.002$ & $2.4\pm1.8$           & $5.1\pm1.5$ \\
\CLf\tablenotemark{b} & $5.40\pm0.15$  & $1.85^{+\infty}_{-0.13}$ & $1.57^{+0.13}_{-0.13}$  & $1.38^{+0.10}_{-0.05}$ & $1.335\pm0.003$ & $9.4\pm6.2$           & $5.7\pm1.6$ \\
\enddata
\tablenotetext{a}{\citet{mwy09}}
\tablenotetext{b}{\citet{wmy09}}
\label{tab_properties}
\end{deluxetable}

\clearpage



\end{document}